\documentclass[aps,prb,twocolumn,letterpaper,showpacs,10pt]{revtex4-1}
\usepackage{graphicx,amsmath,mathrsfs}
\usepackage{amssymb}
\usepackage{amsthm}
\usepackage{bm}
\usepackage{float}
\usepackage{epsfig}

\usepackage{color}

\def\bc{\begin{center}}
\def\ec{\end{center}}

\newcommand{\bs}[1]{\boldsymbol{#1}}

\newcommand{\ket}[1]{\left|#1\right\rangle}

\renewcommand {\ec}{\eta_{\gamma}}

\newcommand{\comment}[1]{}

\def\ket#1{\left|#1\right\rangle}

\begin{document}
\title{Lattice construction of pseudopotential Hamiltonians for Fractional Chern Insulators }
\author{Ching Hua Lee}
\author{Xiao-Liang Qi}
\affiliation{Department of Physics, Stanford University, Stanford, CA 94305, USA}

\date{\today}
\begin{abstract}
Fractional Chern insulators are novel realizations of Fractional Quantum Hall states in lattice systems without orbital magnetic field. These states can be mapped onto conventional fractional quantum Hall states through the Wannier state representation (Phys. Rev. Lett. {\bf 107}, 126803 (2011)). In this paper, we use the Wannier state representation to construct the pseudopotential Hamiltonians for fractional Chern insulators, which are interaction Hamiltonians with certain ideal model wavefunctions as exact ground states. We show that these pseudopotential Hamiltonians can be approximated by short-ranged interactions in fractional Chern insulators, and that their range will be minimized by an optimal gauge choice for the Wannier states. As illustrative examples, we explicitly write down the form of the lowest pseudopotential for several fractional Chern insulator models like the lattice Dirac model and the checkerboard model with Chern number $1$, and the $d$-wave model and the triangular lattice model with Chern number $2$. The proposed pseudopotential Hamiltonians have the $1/3$ Laughlin state as their groundstate when the Chern number $C_1=1$, and a topological nematic $(330)$ state as their groundstate when $C_1=2$. Also included are the results of an interpolation between the d-wave model and two decoupled layers of lattice Dirac models, which explicitly demonstrate the relation between $C_1=2$ fractional Chern insulators and bilayer fractional quantum Hall states. The proposed states can be verified by future numerical works, and in particular provide a model Hamiltonian for the topological nematic states that have not been realized numerically.

\end{abstract}
\maketitle

\section{Introduction}
Chern Insulators (CI) are two-dimensional electron systems which generalize Integer Quantum Hall (IQH) \cite{Klitzing1980} states to band insulators. In CIs, a geometric gauge field is defined in momentum space by the adiabatic transport of Bloch states. The net flux of the gauge field in the Brillouin zone is always quantized in units of $2\pi$ times an integer $C_1$ which is known as the first Chern number\cite{Thouless1982,Haldane1988}. The latter determines the quantized Hall conductance via $\sigma_H=\frac{e^2}{h}C_1$. Soon after the discovery of IQH states, fractional quantum Hall (FQH) states with fractionally quantized Hall conductances were also realized experimentally, with Landau levels partially filled by interacting electrons. Recently, FQH states have also been generalized to lattice systems without orbital magnetic field, which are therefore known as fractional Chern insulators (FCI)\cite{Tang2011,Venderbos2011,Sun2011,Neupert2011}. Fractional Chern insulators are putatively realized in partially filled energy bands with narrow band width (almost flat bands) and commensurate filling, and evidence in support of them have been found in lattice analogs of various FQH states such as Laughlin $1/m$ states, hierarchy states and non-Abelian states\cite{Hu2011,Parameswaran2012,Sheng2011,Tang2011,Wang2011,Wu2012,Goerbig2012,Roy2012,Murthy2011,Murthy2012,Neupert2011a,Neupert2012,Venderbos2012,Sun2011,Kourtis2012,Wu2012a,Jain1989,Liu2013,Lauchli2012}. The situation has been more murky for Chern bands with $C_1>1$, where there is no obvious correspondence between the Chern band and a Landau level whose Chern number is always one\cite{Wang2011a,Trescher2012,Yang2012,Liu2012,Sterdyniak2013,Grushin2012,Ardonne1999}. However, the one-dimensional Wannier state representation (WSR) allows one to systematically understand FCI states based on existing understanding of conventional FQH states by means of an exact mapping between them\cite{Qi2011}. The WSR has been further developed and applied to different FCI states\cite{Barkeshli2012,Wu2012b,Wu2013,Lee2013,Jian2013,Scaffidi2012}, and the ansatz wavefunctions defined by the WSR have been shown to have a high overlap with the exact ground states obtained by exact diagonalization\cite{Wu2012b}.

Besides providing ansatz wavefunctions, the WSR can also be used to construct ideal Hamiltonians for the FCI states. Pseudopotential Hamiltonians\cite{Haldane1983,Moore1991,Read1999} (PPs) have been defined for a large class of FQH states. These are parent hamiltonians which admit the FQH states as unique ground states. Through the WSR, the PP hamiltonians can be mapped onto FCI systems as ideal model Hamiltonians\cite{Qi2011,Lee2013,Wu2013}. These resulting FCI PP Hamiltonians are quasi-local in the sense that the matrix elements decay exponentially in the distance between the sites involved. In our previous work Ref. \onlinecite{Lee2013}, we studied the overlap between the PP Hamiltonian and local density-density interactions in some FCI models, hence obtaining some guidance about the forms of interaction Hamiltonians that may lead to FCI phases. We also generalized the PP Hamiltonian approach from the two-body case to the many-body case. While the study of overlaps can tell us how promising the interactions are in stabilizing topologically nontrivial groundstates, it does not tell us how to do better - what will be the {\it optimal} kind of interaction for the realization of a given state? That will be the goal of this paper.

In this paper, we construct the ideal Hamiltonians for FCI's by reverse-engineering the lattice Hamiltonians from the PP Hamiltonians. This approach has the advantage of producing "ideal" interaction operators that by construction are guaranteed to possess certain grounstates. By starting directly from these operators, we will not be plagued by situations where a significant part of a given interaction cannot be expanded in terms of the PPs, a consequence of the incompleteness of the space of PPs\cite{Lee2013}. We first write the PP Hamiltonians for FQH states in the Landau gauge, and then map the Landau gauge wavefunctions to the Wannier states of the lattice system using the WSR. This procedure defines the lattice PP Hamiltonian, which is a quartic term in the boson or fermion creation operators when the original FQH pseudopotential involves two bodies. A Chern band with Chern number $C_1$ is mapped to $C_1$ ``layers" of Landau levels, and the corresponding PP Hamiltonian is obtained from that of a $C_1$ layer FQH system.The PP Hamiltonian obtained in this way is not completely local, although all long range terms have exponentially small weights. In the final step of this procedure, we truncate the lattice Hamiltonian by neglecting all terms beyond a certain range, and obtain an interaction Hamiltonian that is the best approximation of the the corresponding PP Hamiltonian among other interaction terms with a similar range.

Model interaction operators obtained from the above procedure generically suffer from a proliferation of density-density and hopping terms. Hence, a principal aim of this paper is to investigate how the various degrees of freedom in the ambiguities of specifying the PPs - to be discussed later - can be exploited to produce PPs dominated by the least number of terms. One ambiguity is inherent in the WSR, where the Landau gauge wavefunctions of the lowest Landau level (LLL) in the QH system is mapped to the 1D Wannier states of the FCI system. Since the 1D Wannier states are defined by a fourier transformation of the Bloch states, there is a gauge ambiguity in the definition of the Bloch states. In Ref. \onlinecite{Qi2011}, the gauge fixing was done by making the Wannier states maximally localized. Since then, different gauge choices have been studied to optimize the ansatz wavefunction\cite{Wu2012b,Wu2013}. The point group symmetries of the lattice Hamiltonian can also impose constraints to the gauge choices\cite{Jian2013}. We hence develop a new gauge fixing scheme that produces PPs with maximal locality. The range of PP Hamiltonians can be determined from that of a lattice coherent state obtained from mapping the Landau level coherent states (Gaussian wavepackets) to the FCI system. Different gauge choices lead to coherent states of different shapes. By requiring the coherent state to be maximally localized, we determine the optimal gauge choice which coincides with the Coulomb gauge proposed in Ref. \onlinecite{Wu2013}. This gauge choice is different from the gauge that maximally localizes the WFs which was used in our previous work\cite{Lee2013}, although the difference is usually small and vanishes entirely in the limit of uniform Berry curvature. Interestingly, certain bounds on the locality are uniquely determined by the geometric properties of Bloch wavefunctions, {\it i.e.} the Berry's curvature and the Fubini-Study metric.

As illustrative examples, we provide numerical results on the form of the lowest PPs in various FCI systems studied in literature. For $C_1=1$, we study the commonly used Checkerboard (CB) flatband model\cite{Sun2011} and the lattice Dirac model. In particular, we found that the first PP of the latter is dominated by a single nearest-neighbor interaction term. For $C_1=2$ we studied the triangular lattice model used in Ref. \onlinecite{Wang2011} and a square lattice model with hopping terms of $d$-wave symmetry. To illustrate the stability of the leading contributions in the PP operator, we also present an interpolation between the $d$-wave model and two decoupled layers of lattice Dirac model. The interpolation also explicitly illustrate the relation between $C_1=2$ FCI and bilayer FQH states proposed in Ref. \onlinecite{Barkeshli2012}, and provide the first model Hamiltonian for the $(mm0)$ topological nematic state with non-Abelian lattice dislocations.

The rest of the paper is organized as follows. In Sec. \ref{sec:preliminaries} we review the framework of WSR and the construction of the PPs using a coherent state basis. In Sec. \ref{sec:MLCS} we find the maximally localized coherent states by optimizing the gauge choice of the Bloch states, which are then used to construct PPs with shortest range. In Sec. \ref{sec:numerics} we present the numerical results of truncated pseudopotential Hamiltonians in several FCI models. Finally, Sec. \ref{sec:concl} is dedicated to the conclusion and discussion of future works.

\section{Preliminaries}\label{sec:preliminaries}

In this section, we shall review the pseudopotential Hamiltonian construction in the Landau level problem and its FCI generalization. Within a Landau level of the QH system, the kinetic energy is frozen out and the Hamiltonian only contains an interaction term

\begin{equation}
H=\sum_{i<j} V (\bs{r}_i-\bs{r}_j),
\end{equation}
where the sum extends over all pairs of particles. Here and below, it should be remembered that the Hamiltonian is projected to states in a Landau level, although for simplicity we will not write the projection explicitly.  Translation symmetry is assumed, so that interaction only depends on the relative position ${\bf r}_i-{\bf r}_j$.
The two-body interaction Hamiltonian $H$ can be decomposed into relative angular momentum of the two particles:
\begin{eqnarray}
H&=&\sum_{i<j} \sum_{m=0}^{\infty} V^m P^m_{ij},\\
P^m_{ij}&=&L_m (-l_B^2 \nabla^2) \delta^2 (\bs{r}_i -\bs{r}_j)\nonumber
\end{eqnarray}
where $L_m$ are the Laguerre polynomials\footnote{for a derivation of this expression and an introduction to PPs, see \onlinecite{Lee2013} and references therein.}. $ P^m_{ij}$ projects onto a state where particles $i$ and $j$ have relative angular momentum $m$ and $V^m$ is the energy penalty
for having two particles in such a state. When the particles are bosons(fermions), only terms with even(odd) $m$ need to be considered. As a component with fixed relative angular mometnum,the two-body interaction potential $P^m_{ij}$ is known as a pseudopotential (PP)\cite{Haldane1983,Trugman1985}. We denote the prototypical PPs representing interactions with only one nonzero $V^m$ by $U^m$:
\begin{equation}
U^m=  \sum_{i<j} P^m_{ij}
\label{lag}
\end{equation}
The Laughlin $1/m$ state lies in the nullspace of $U^n$ for all $n<m$. Hence it will be unique groudstate for hamiltonians with positive $n^{th}$ PP coefficients for all $n<m$, i.e. the Laughlin state at $1/5$ filling is the unique grounstate of $U^1+U^3$.

This paper will be primarily about the mapping of the PP interactions $U^m$ to generic FCI systems. As explained in the introduction, relative angular momentum is ill-defined on the lattice. However, the PP's $U^m$ are still well-defined via a correspondence between the basis of the FCI and QH systems. Specifically, we introduce an unitary mapping between the Hilbert spaces of the QH and the FCI systems \cite{Qi2011,Lee2013}:
\begin{eqnarray}
f:~{\rm H}_{\text{FCI}}&\longrightarrow& {\rm H}_{\text{QH}},\nonumber\\
f\left(\ket{W_K}\right)&=&\ket{\psi_K},
\label{mapFQHFCI}
\end{eqnarray}

with ${\rm H}_{\text{FCI}}$ and ${\rm H}_{\text{QH}}$ denoting the Hilbert spaces of the FCI and QH systems respectively. At the operator level, we have $f\left(b^\dagger_K\right)= a^\dagger_K$ where $a^\dagger_K\ket{0}=\ket{\psi_K}$ and $b^\dagger_K\ket{0}=\ket{W_K}$. Here $\{|\psi_K\rangle\}$ are the Laudau gauge wavefunctions which span the Hilbert space of the LLL in the QH system, while $\{|W_K\rangle\}$ spans the occupied states of the FCI system.

Expressed in real-space, the Landau gauge wavefunctions of the QH system takes the form

\begin{equation} \psi_{K=\frac{2\pi n }{L}}(x,y)=\sum_{K'\in \mathbb{Z}}\frac{1}{\sqrt{\sqrt{\pi}Ll_B}}e^{i(K+K')y}e^{-A(\frac{x}{l_B}-l_B(K+K'))^2}
\label{LLLbasis}
\end{equation}

The sum in the RHS enforces the toroidal boundary conditions of the system. Here $A$ is an aspect ratio parameter that can be tuned through rescalings of the coordinates. When the rescaling is regarded as an active transformation, it can formally be represented by a redefinition of the intrinsic QH metric\cite{Haldane2011,Yang2012a,Yang2012b}\footnote{We write the FQH Hamiltonian as $ H=\frac{1}{2m}g^{ab}\pi_a\pi_b $ where $\pi_a=p_a-\frac{e}{c}A_a(\vec r)$. $g^{ab}$ represents an intrinsic QH metric that reduces to the identity $\delta^{ab}$ in the usual isotropic case. Expressed a quadratic form, $g^{ab}$ specifies the eccentricity and orientation of the generically elliptical LL orbital in the symmetric gauge. For simplicity, we will consider the case of zero orientation shift, so that $g=diag(C_1/A,A/C_1)$ where $C_1$ is the Chern number and $A$ is an aspect ratio.}. When regarded as a passive transformation, it can be  interpreted as a LL basis redefinition which elegantly corresponds to Bogoliubov transformations of the QH second-quantized operators in the symmetric gauge\cite{Qiu2012}.

In the FCI system, the Wannier function $W_K$ is given by a linear superpositions of the periodic part of the Bloch states $\phi(k_x,k_y)$:
\begin{equation}
W_{K=k_y+2\pi x}(x,y)=\int dk_x e^{i(k_x x+ k_y y)}e^{i\theta(k_x,k_y)}\phi(k_x,k_y)
\label{wannierdef}
\end{equation}
with the gauge phase $e^{i\theta}$ arising from the intrinsic arbitrariness of the phase of the Bloch states. $\ket{W_K}$ is taken to be the analog of the LLL Landau gauge wavefunction $\ket{\psi_K}$ of the QH system because such a mapping preserves the continuity in $K$ and also the topological properties of $\ket{W_K}$ and $\ket{\psi_K}$, i.e., the centers-of-mass of both shift by one site as $K$ (or $k_y$) is translated by $2\pi$.\cite{Qi2011,Lee2013,Thonhauser2006} This is analogous to the behavior of the $\psi_K$, as can be seen from its gaussian factor $e^{-\frac{A}{2l_B^2}(x-l_B^2K)^2}$. Hence the analogy between these two bases is completed by setting the effective magnetic length to be $l_B=\sqrt{\frac{1}{2\pi}}$ in units of the FCI lattice spacing.

Since we want to maximize the locality of the $U^m$ operators in the FCI system, we shall first express them in a form where their real-space locality can be directly studied. This can be explicitly done in the QH case, after which the FCI case follows by simply replacing $a^\dagger_K$ with the WF creation operator $b^\dagger_K$, as mentioned below Eq. \ref{mapFQHFCI}. As shown in Appendix \ref{ppcoherent}, one can use Eqs. \ref{lag} and \ref{LLLbasis} to bring QH pseudopotential $U^m$ to the useful form involving coherent states:\cite{Qi2011,Lee2013}

\begin{equation}
U^m\propto \int dz_1dz_2 \sum_r v^m_r \nabla_z^m (c^\dagger_z c_z ) \nabla_z^m (c^\dagger_z c_z )
\label{main}
\end{equation}
where $v^m_r$ are the coefficients of the Laguerre polynomials $ L_m(x)=\sum^{m}_{r=0} v^m_r x^r $, $z=z_1+iz_2$, and $\nabla_z$ is to be taken as a grad operator so that $\nabla_z^3 f=\nabla_z (\nabla_z\cdot \nabla_z f)$, etc.

The coherent state creation and annihilation operators are defined by\cite{Qi2011}:
\begin{equation}
c^\dagger_z =  \sum_{K} e^{-iz_2K}e^{-\pi A(z_1-K/(2\pi))^2}a^\dagger_K
\label{coherentdef}
\end{equation}
As will be made evident in the next section, this way of expressing $U^m$ makes its locality explicit because the coherent state created by $c_z^\dagger$ has a local wavefunction in real space (in the sense of exponential decay which we will demonstrate shortly after). Hence our attempt to maximally localize the PPs is reduced to a search for a basis with the smallest coherent state spread.

For simplicity, we shall first consider the case with Chern number $C_1=1$ in the lattice system, which is a direct analog of a QH system with one LL. The coherent state $\ket{\Psi_z}=c^\dagger_z\ket{0}$ of the lattice system can be written, via the replacement $a^\dagger_K\rightarrow b^\dagger_K$, in the following convenient form

\begin{eqnarray}
&&\Psi_z(x,y)\notag\\
&=&\sum_K e^{-ik_yz_2}e^{-\pi A(z_1-K/{2\pi})^2}W_K(x,y)\notag\\
&=&\sum_m\int d^2k e^{-ik_yz_2}e^{i\vec k \cdot \vec r}e^{-\pi A\left(z_1-(\frac{k_y}{2\pi}+m)\right)^2}e^{i\theta}\phi(k_x,k_y)\notag\\
\label{coherentformc1}
\end{eqnarray}
with $\vec{r}=(x,y), ~\vec{k}=(k_x,k_y)$ here and below. That $\Psi_z(x,y)$ is asymptotically exponentially decaying in $x,y$ follows from the fact that $\phi(k_x,k_y)$ in the integrand has a complex singularity at a finite distance from the real $k_x$ or $k_y$ axis \footnote{More generally, the fourier coefficients of a complex function decay exponentially at a rate $h$ whenever the function has a singularity at a distance $h$ from the real axis. This is proven in Ref. \onlinecite{he2001}, and also in C.H.Lee and R.Thomale 2014 (in progress)}.

For lattice systems with general $C_1$, there will be $C_1$ coherent states "colors" in a multiplet\cite{Barkeshli2012,Wu2013}. This is because each $W_K$ will now shift by $C_1$ sites as $K$ (or $k_y$) is translated by $2\pi$, leading to $C_1$ inequivalent "layers". Indeed, we now have a map from a Chern number $C_1$ lattice system to a multilayer $QH$ system, with each layer having Chern number one and $l_B=\sqrt{\frac{C_1}{2\pi}}$.

For each layer (color), the expression for the coherent state is generalized to
\begin{eqnarray}
&&\Psi_z(x,y)\notag\\
&=&\sum_m\int d^2k e^{-ik_yz_2}e^{i\vec k \cdot \vec r}e^{-\pi A\left(\frac{z_1}{C_1}-(\frac{k_y}{2\pi}+m)\right)^2}e^{i\theta}\phi(k_x,k_y)\notag\\
\label{coherentform}
\end{eqnarray}
which is manifestly expressed in terms of three tunable inputs: The aspect ratio $A$, the gauge phase $\theta$ and the Bloch states $\phi$. To minimize the proliferation of terms in the resultant PP $U^m$, we will need to maximize the locality of the coherent states, as evident from the relation Eq. \ref{main}. The optimal $A$ and $\theta$ will be derived in great detail in the next section.

Note that Eq. \ref{coherentform} reduces to a 2-dimensional fourier transform of the Bloch wavefunction in the absence of the two terms containing $z$. When $C_1\neq 0$, such a fourier transform cannot result in a localized state due to topological obstruction\cite{Thonhauser2006}. As such, the Gaussian term containing $z_2$ can also be regarded as a regularizing factor.

From Eq. \ref{coherentform}, it is evident that we can, by redefining the unit cell, change the bloch states and hence significantly change the form of the coherent states $\Psi_z(x,y)$ and the locality of the PPs. Let the original position coordinates and crystal momentum be labeled as $r=(x,y)$ and $k=(k_x,k_y)$. Under a redefinition of unit cell by a 2-by-2 matrix $M$,

\begin{eqnarray} r\rightarrow r'&=&Mr \\
 k\rightarrow k'&=&M^{-1}k \end{eqnarray}

with the scalar product $k\cdot r$ staying invariant. Under general transformations $M$, $\Psi_z(x,y)$ given by Eq. \ref{coherentform} will change unless $k_y'=k_y$ and the Bloch states $\phi(k)$ remain invariant under $M$. Manifest here is the asymmetry between the roles played by $k_x$ and $k_y$, an unavoidable feature of the nature of the Landau gauge wavefunctions and their analogs.

There is a special class of transformations $M$ where all the terms in the definition of $\Psi_z$ Eq. \ref{coherentform} remain invariant, except for the Bloch state $\phi(k)\rightarrow \phi(k')$. In other words, the PPs will be exactly that obtained for the same Hamiltonian with $k\rightarrow k'$, up to the rescaling of coordinates $r=(x,y)$. These transformations are given by

\begin{eqnarray}
x'&=&\alpha x \notag\\
y'&=&y-\beta x
\label{abredef1}
\end{eqnarray}

\begin{eqnarray}
k'_x&=&\frac{k_x+\beta k_y}{\alpha} \notag\\
k'_y&=&k_y
\label{abredef2}
\end{eqnarray}
We will revisit transformations of this type in Sec. \ref{sec:numerics}, where we compare the PPs of the same Dirac model after some coordinate redefinitions.

Before further investigations on the pseudopotential in Eq. \ref{main}, we will like to discuss its symmetry properties. Although the Chern band is topologically equivalent to a Landau level system, the latter possess a higher symmetry which includes all symmetries of the Chern band. More specifically, the pseudopotentials $U^m$ in FQH systems preserve the magnetic translation symmetry which which is a higher symmetry than the lattice translation symmetry of the lattice. In a FQH system spanned by the Landau gauge basis $\ket{\psi_K}$, the magnetic translation symmetry implies the translation symmetry of $K$ to $K+\frac{2\pi}L$. However, a generic Chern band does not have this symmetry due to its non-uniform Berry's curvature. This non-uniformity is clearly manifested in the nonlinearity of the center-of-mass of the Wannier basis, as detailed in Ref. \onlinecite{Qi2011}. Therefore the psedopotentials $U^m$ that we defined in FCIs have a higher symmetry than a generic interaction term preserving the lattice symmetries. This is not a concern for our current work since our purpose is to explicitly write down pseudopotential interactions which can then be used in numerical studies for realizing certain topological states as ground states. 

An alternative explicit expression for $U^m$ have also been derived in Ref. \onlinecite{Lee2013}, where they take an elegant form in a one-dimensional basis proposed in Ref. \onlinecite{Lee2004}. While that form allows for easy comparisons between different interactions at the operator level, the forms derived above (i.e. Eqs \ref{main},\ref{coherentform}) are more suitable for locality optimization, being explicitly defined in position space.

\section{Maximally localized coherent states}
\label{sec:MLCS}
\subsection{The optimal gauge phase}

In this section, we show how to minimize the spread of the coherent states $\Psi_z$ so that the PPs will be most localized and be best approximated by a short-ranged interaction. For this purpose, we define the mean-squared range of the coherent state wavefunction $\Psi_z(x,y)$ by
\begin{eqnarray}
&&I[\theta]=\int_{-\frac{1}{2}}^{\frac{1}{2}}\int_{-\frac{1}{2}}^{\frac{1}{2}} d^2z\langle r^2\rangle_{\Psi_z}\nonumber\\
&=&\int_{-\frac{1}{2}}^{\frac{1}{2}}\int_{-\frac{1}{2}}^{\frac{1}{2}}d^2z \int dxdy (x^2+y^2) |\Psi_z(x,y)|^2 \nonumber \\
&=&\int_{-\frac{1}{2}}^{\frac{1}{2}}\int_{-\frac{1}{2}}^{\frac{1}{2}}d^2z \sum_m \int d^2k \left|\nabla_k \left (e^{-E_m-ik_yz_2}e^{i\theta }\phi(k_x,k_y)\right )\right |^2\nonumber\\
\label{coherentI}
\end{eqnarray}
which is regarded as a functional over the $U(1)$ gauge phase $\theta(k_x,k_y)$. Here $E_m(k_x,k_y)=-\pi A \left(\frac{k_y}{2\pi}-\frac{z_1}{C_1}+m\right )^2$. For simplicity, we will only consider the case with one occupied band.
Since we fourier transformed over the concatenated momentum $K= k_y+ 2\pi m$ from the penultimate to the last line, the $m$ sum must be included. $A$ represents the QH aspect ratio that is yet to be optimized.

We had sought to minimize the spread $\langle r^2\rangle$ of $\Psi_z$ averaged for different $z$. This averaging is motivated by various reasons. Firstly, the optimal localization condition for each $\Psi_z$ will lead to a different condition on gauge choice $\theta$, and we desire for an unique optimal gauge choice. Secondly, it makes physical sense to optimize the average spread since we will require the locality of the coherent states over all $z$ when calculating the PP. Furthermore, performing the average over $z$ will greatly simplify the minimization problem, as we will soon see. As distinguished from the FQH case, in FCI the coherent states with different center-of-mass position $z$ generically have different shapes, since the translation symmetry is broken to discrete lattice translation. Since lattice translation symmetry guarantees that $\Psi_z$ has the same spread as $\Psi_{z+1}$ and $\Psi_{z+i}$, we will only need to average across the unit cell $(z_1,z_2)\in\left[-\frac12,\frac12\right]\times \left[-\frac12,\frac12\right]$.

Given a Bloch wavefunction $\phi$ with Berry connection $\vec{a}=-i\phi^\dagger \nabla_k \phi$, a gauge rotation $\phi\rightarrow e^{i\theta}\phi$ will produce a new connection $\vec{a}\rightarrow \vec{a}_{new}=\vec{a} + \nabla_k\theta$.
Performing the Euler-Lagrange minimization \[\frac{\delta I}{\delta \theta}=0,\] whose detailed steps are shown in Appendix \ref{gaugederivation}, we arrive at the \textbf{Coulomb gauge condition}
\begin{equation}
\nabla_k \cdot\vec{a}_{new}=0
\label{eq:coulomb}
\end{equation}
We note that Eq.~\ref{eq:coulomb} coincide with the gauge choice made in Ref. \onlinecite{Wu2013}, although it was not obtained by maximal localization condition over there.

In this gauge choice, there exists a simple way to express $\vec{a}_{new}$ in terms of the Berry curvature $f$. This gauge is also consistent with the conditions for a $C_4$ symmetric basis in the case of $C_4$ symmetric systems, as shown in Ref. \onlinecite{Jian2013}. Since $\nabla_k\cdot\vec{a}_{new}=0$, we can write $\vec{a}_{new}=(-\partial_y\varphi,\partial_x \varphi)^T$, where

\begin{equation}\nabla_k^2 \varphi(k_x,k_y) = f(k_x,k_y)=\partial_xa_y-\partial_ya_x
\label{eq:poisson1}
\end{equation}

This is the Poisson's equation on a finite torus whose explicit solution is given by Eq. $9$ of Ref. \onlinecite{Wu2013}. When the dimensions of the torus $L_x,L_y>10$, as is the case for the calculations in this work, the above solution will be almost identical to that given by the electrostatics Green's function integral

\begin{equation} \varphi (k_x,k_y)= \int_{periodic} f(\vec{p})\log|\vec{p}-\vec{k}| d^2p
\label{eq:poisson2}
\end{equation}
where the contributions from the periodic images of $f$ are summed over. Here $\varphi$ and $f$ take the role of the electrical potential and the periodic electric charge distribution.

To calculate the PPs, however, we need to find the gauge phase $\theta$ and not just $\vec{a}_{new}=(-\partial_y\varphi,\partial_x \varphi)^T$. The optimal gauge phase $\theta=\theta_C$ can be found from the condition Eq. \ref{eq:coulomb}, which upon subtituting $\vec{a}_{new}=\vec{a} + \theta_C$ gives another Poisson's equation

\begin{equation}\nabla_k^2\theta_C =-\nabla_k\cdot \vec{a} \end{equation}
The solution to this equation is already given by Eq.~\ref{eq:poisson2} with $\varphi$ and $f$ replaced by $\theta_C$ and $\nabla_k \cdot \vec{a}$. Note that the solution $\theta$ obtained is unique, because the difference of two different solutions will satisfy the Laplace equation, and that must be identically zero due to toroidal BCs.

There is a nice relation between Coulomb gauge $\theta_C$ (obtained by solving Eq.~\ref{eq:coulomb}) which maximally localizes the coherent states and $\theta_{W}$, the gauge that maximally localizes WFs. The latter, which was used in our previous work Ref. \onlinecite{Qi2011}, is given by

\begin{equation}
\theta_{W}(k_x,k_y)=-\int_0^{k_x} a_x(p_x,k_y)dp_x +\frac{k_x}{2\pi}\int_0^{2\pi} a_x(p_x,k_y)dp_x
\label{eq:XL}
\end{equation}

\begin{figure}[H]
\begin{minipage}{0.99\linewidth}
\includegraphics[width=.5\linewidth]{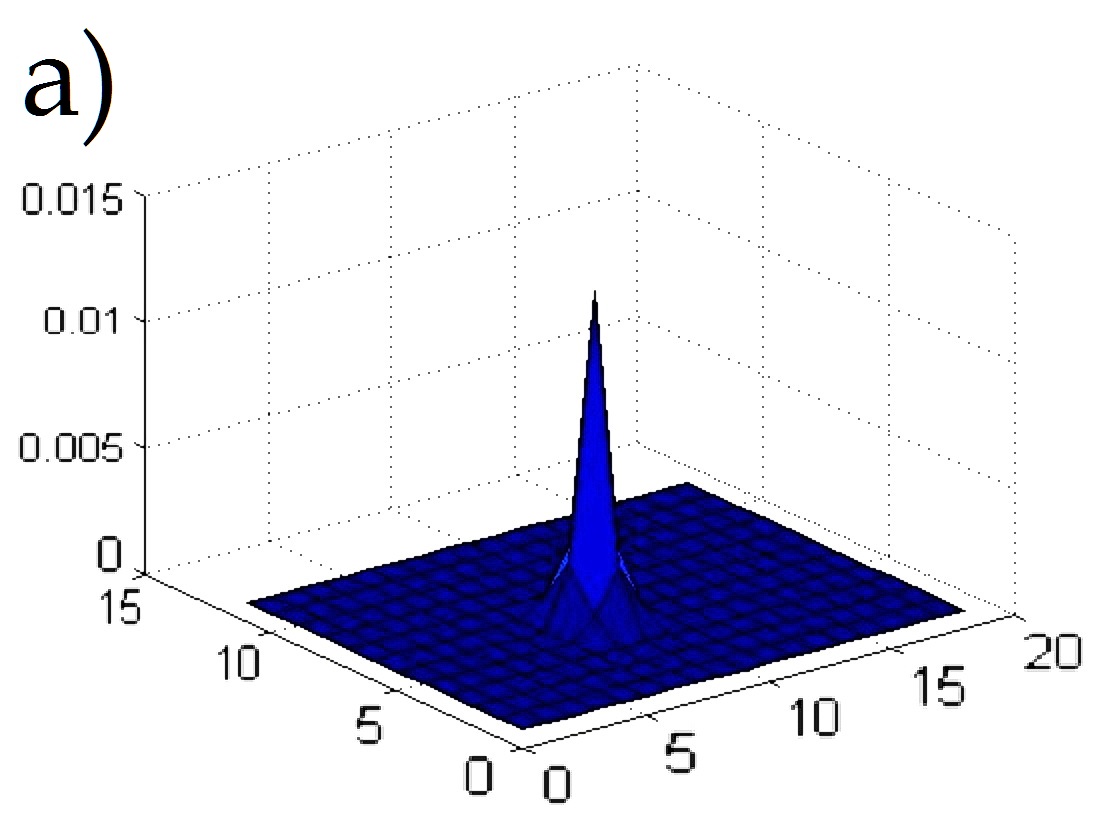}
\includegraphics[width=.5\linewidth]{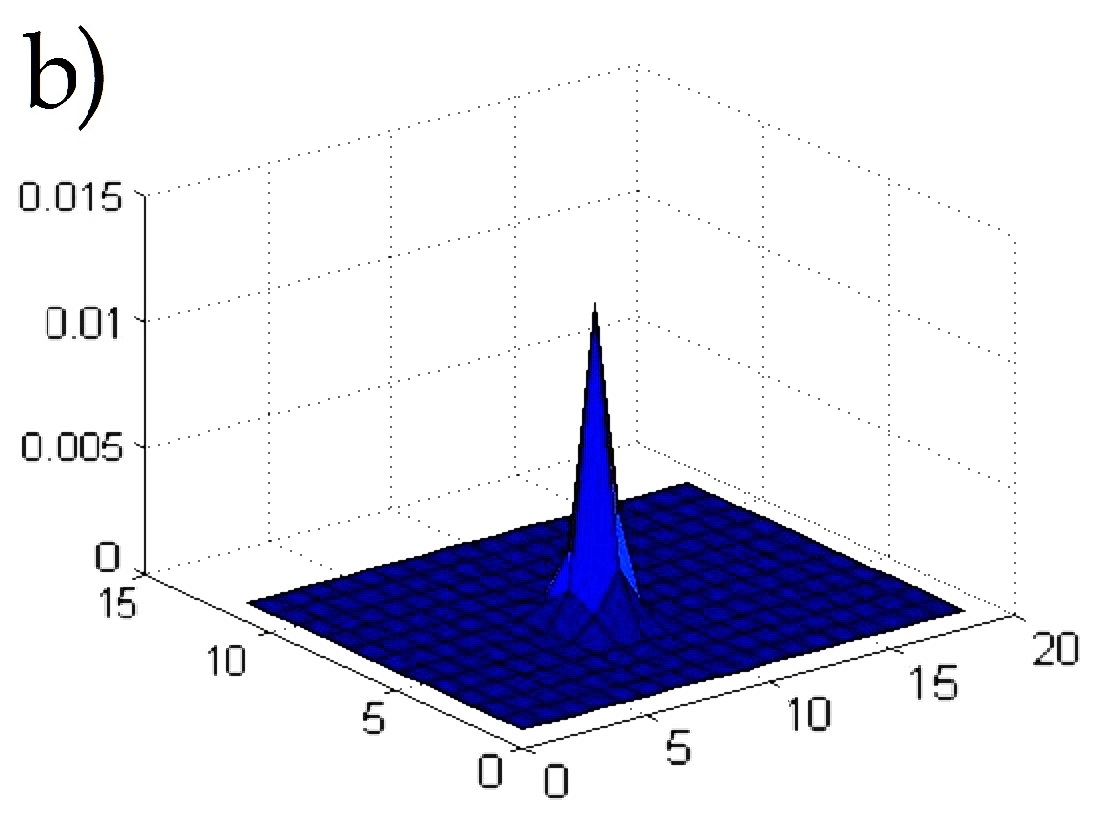}
\end{minipage}
\caption{(Color Online) A typical coherent state $|\Psi_z|^2$ of the $C_1=2$ d-wave model (see Eq. \ref{interpol1}) computed with the gauge phase that maximizes the localization of (a) the Wannier states and (b) the coherent states.
We see that there is very little difference in their spread, as expected from the relatively flat gauge curvature. }
\label{fig:twogauges}
\end{figure}

We shall show that these two gauge phases differ only by a relatively small correction, at least for most sufficiently smooth Berry curvatures. Write

\[\theta_C= \theta_{W}+\theta_0\]

where $\theta_C$ is the Coulomb gauge phase and $\theta_0$ is a relatively small correction. We have

\begin{eqnarray}
&&\nabla_k^2\theta_0(k_x,k_y)\nonumber\\
&=& -\partial_x a_x -\left(-\partial_x a_x -\int_0^{k_x} \partial^2_y a_x dp_x +\frac{k_x}{2\pi}\int_0^{2\pi}\partial^2_y a_x dp_x \right)\nonumber \\
&=&\partial_y\left(\int_0^{k_x} f(p_x,k_y)dp_x -\frac{k_x}{2\pi}\int_0^{2\pi}f(p_x,k_y)dp_x \right),
\label{fy}
\end{eqnarray}
so that
\begin{equation} \nabla_k^2 (\partial_x\theta_0(k_x,k_y))= \partial_y (f(k_x,k_y)-\bar f (k_y))\end{equation}
where the Berry curvature $f=\partial_y a_x$ and $\bar f(k_y)=\frac1{2\pi}\int dk_x f(k_x,k_y)$ is $f$ averaged over $k_x$. Hence the correction $\theta_0$ is directly related to the nonuniformity of the Berry curvature, and disappears when the latter is constant w.r.t. to $k_x$ or $k_y$ (or both). We should thus expect it to be relatively small in magnitude. The comparison between the two gauge choices is shown in Fig. \ref{fig:twogauges}.

\subsection{Properties of the average coherent state spread}

Interestingly, the average spread of the coherent state $\int_{[-1/2,1/2]^2} d^2z\langle r^2\rangle_{\Psi_z}$ can be expressed in terms of the geometric quantity $\Omega_1$, to be defined shortly, and the Berry curvature $f$. As shown in Appendix \ref{spread}, the average spread of the coherent state can be rewritten as

\begin{eqnarray}
I[A]&=&\int_{[-1/2,1/2]^2} d^2z\langle r^2\rangle_{\Psi_z}\nonumber \\
&=&\frac{1}{\sqrt{2A}}\int d^2k \left(\left[|\nabla_k \phi|^2-|\vec{a}|^2\right]   + |\vec{a}_{new}|^2+\frac{1}{12} +\frac{A}{4\pi}\right) \nonumber \\
\label{eq:spread}
\end{eqnarray}
where $\vec{a}_{new}=\vec{a}+\nabla_k\theta_C$ is the Coulomb gauge connection that will minimize the spread. Note that $I[A]$ depends on $A$, the FQH aspect ratio that has yet to be chosen. Its optimal value will be the subject of the next subsection; here we will explore its $A$-independent properties.

Of central interest is the quantity $J$ given by

\begin{eqnarray}
J&=&\int d^2k\left(|\nabla_k \phi|^2-|\vec{a}|^2\right)   + |\vec{a}_{new}|^2
\label{J}
\end{eqnarray}

This quantity can be expressed in terms of the geometrical properties of the system. First, consider the bracketed terms

\begin{equation}
\Omega_1=\int d^2k \left(|\nabla_k \phi|^2-|\vec{a}|^2\right)
\label{omega1eq}
\end{equation}

$\Omega_1$ is actually a gauge invariant. This can be seen directly by replacing $\phi$ by $e^{i\theta}\phi$, or by rewriting $\Omega_1$ as a trace over certain position operators projected onto the occupied states. This will be shown in much more detail in Appendix \ref{omega1}, where its connection with the Wannier operator will also be made more explicit.

$\Omega_1$ can also be expressed in terms of the Fubini-Study metric $g$ associated with the quantum distance between two points on the BZ, i.e. $ds^2_{12}=1-|\langle \phi_1|\phi_2\rangle|^2$. Here $g_{ab}(k)=Re\langle \partial_a \phi|\partial_b \phi\rangle - \langle \partial_a \phi|\phi\rangle\langle\phi|\partial_b \phi\rangle $ (See Ref. \onlinecite{Marzari1997} for a multiband expression). Referring back to Eq. \ref{omega1eq}, we see that

\begin{equation}
\Omega_1 = \frac{1}{4\pi^2}\int d^2k Tr g
\end{equation}

i.e. $\Omega_1$ is a trace over the metric integrated over the BZ. This is a geometrical quantity, and is obviously gauge-invariant. Physically, it is a measure of how orthogonal the Bloch states on neighboring points the BZ are: a large contribution to the integral arises at regions of the BZ where even nearby states quickly become orthogonal
, i.e. $ds^2\rightarrow 1$.

The last term of $J$ as shown in Eq. \ref{J} is geometrical too, albeit not gauge-invariant. It can be expressed in terms of the Berry curvature of the system as we shall see below. In the optimal Coulomb gauge derived previously, $\nabla_k\cdot\vec{a}_{new}=0$. This permits us to write $\vec{a}_{new}=(-\partial_y\varphi,\partial_x \varphi)^T$, where

\[\nabla_k^2 \varphi(k_x,k_y) = f(k_x,k_y), \]

Hence $J$ can be written in terms of manifestly geometric quantities as follows:

\begin{eqnarray}
J&=&\int d^2k\left( [|\nabla_k \phi|^2-|\vec{a}|^2] + |\vec{a}_{new}|^2\right)\notag\\
&=& \Omega_1  + \int d^2k \left(|\partial_x \varphi|^2+|\partial_y \varphi|^2\right )\notag \\
&=& \Omega_1 - \int d^2k \varphi \nabla_k^2 \varphi =\Omega_1 - \int d^2k\varphi f   \notag \\
&=& \int d^2k \left [\frac{1}{4\pi^2}Tr g - \varphi f  \right]
\end{eqnarray}

On the last line, $\varphi$ is derived from the geometric Berry curvature $f$ via $\nabla_k^2\varphi = f$. Hence the optimal average spread in Eq. \ref{eq:spread} can be computed wholly from the Fubini-Study metric and the Berry Curvature of the system without first computing the Coulomb gauge connection.

Since we have specialized to the Coulomb gauge, $J$ can also be simplified via

\begin{eqnarray}
J&=& \int d^2k \left(|\nabla_k\phi_{new}|^2-|\vec{a}_{new}|^2\right)   + |\vec{a}_{new}|^2\notag \\
&=&\int d^2k |\nabla_k \phi_{new}|^2
\end{eqnarray}

where $\phi_{new}$ is the Bloch wavefunction in the Coulomb gauge. $J$ superficially looks like a real-space spread $\langle r^2\rangle$ of $\phi_{new}$. However, this cannot be strictly true, since for our regime of interest $C_1\neq 0$, $\phi_{new}$ is not periodic in the BZ and $\nabla_k^2$ cannot be fourier transformed into $r^2$. In fact, our coherent Wannier state construction circumvents this difficulty by involving Gaussian-weighted concatenated Wannier states that are periodic\footnote{If $C_1=0$, however, $\phi_{new}$ can be made periodic with a suitable choice of gauge and $J$ does indeed represent its real-space spread. In this case, $I[A]$ will be irrelevant as the coherent-state construction will be unnecessary.}. Note that $J$ is fundamentally different from $I[A]$, because the latter involves an averaging of coherent states over different centers-of-mass while the former involves only the Bloch states.

\subsection{The optimal aspect ratio $A$}

Setting $\frac{dI}{dA}=0$ in Eq.~\ref{eq:spread}, we obtain the optimal value of the aspect ratio $A$ as

\begin{eqnarray}
A_{opt}= \frac{1}{\pi}\int d^2k \left(\left[|\nabla_k \phi|^2-|\vec{a}|^2\right]   + |\vec{a}_{new}|^2+\frac{1}{12} \right) =  \frac{J}{\pi} + \frac{\pi}{3}\notag\\
\label{eq:aopt}
\end{eqnarray}
where $ J = \Omega_1 - \int d^2k\varphi f 
= \int d^2k |\nabla_k \phi_{new}|^2$.

For this optimal aspect ratio, the gaussian factor in the coherent state wavefunction (\ref{coherentformc1}) reads

\begin{eqnarray}
&&-\pi A\left(\frac{z_1}{C_1}-\frac{2\pi m +k_y}{2\pi}\right)^2\nonumber \\
&=&-\frac{1}{C_1^2}\left(J+\frac{\pi^2}{3}\right)\left(z_1-C_1\frac{2\pi m +k_y}{2\pi}\right)^2
\end{eqnarray}

Note that the $\frac{\pi^2}{3}$ term comes from the $dz_2$ integration, and can roughly speaking be regarded as a feature of the two-dimensionality of the coherent states. The above result agrees with the numerically obtained typical optimal $A$s of $\approx 2$ when $C=2$ and $\approx 1.4$ when $C=1$.

Substituting this optimal value of $A$ into the spread function $I$, we finally arrive at

\begin{equation}
I_{opt}=\int_{[-1/2,1/2]^2} d^2z\langle r^2\rangle_{\Psi_z}=\sqrt{2\pi \left(J+\frac{\pi^2}{3}\right)}
\end{equation}

As mentioned in the previous section, we can also optimize the locality of the PPs changing the Bloch basis themselves through coordinate redefinitions.
Now, we explicitly see how the locality depends on the smoothness of the Bloch states through
\begin{equation} J = \int d^2k |\nabla_k \phi_{new}|^2. \label{j}\end{equation}.

In the numerical results that follow, we see that systems with more bands generically have more complicated $\phi_{new}$ and hence poor locality of the PPs. This seems to be true for band insulators, although not for the QH LLs which always takes a similar functional form.

\section{Truncated Pseudopotential Hamiltonians: Numerical results}
\label{sec:numerics}

Even after the abovementioned optimization procedure, the wavefunction of a coherent state still has an gaussian decaying tail, which means the PP Hamiltonian (\ref{main}) still contains infinite number of terms when we expand it in the real space basis. To obtain a physical lattice Hamiltonian with finite interaction range, we can take a truncation and keep only a small number of dominant short-range terms. We have numerically studied the first PP Hamiltonian $U^1$ for four different models with Chern bands. The truncated PP Hamiltonians are summarized in Table \ref{table1}, and the definition and more details of these four models are presented in the rest of this section.

\begin{widetext}
\begin{table}[H]
\centering
\renewcommand{\arraystretch}{2}
\begin{tabular}{|l|l|l|}\hline
\textbf{Model} &\ $C_1$ &\ \textbf{The truncated first PP hamiltonian $U^1$ } \\ \hline
Checkerboard  &\ 1 &\ $\displaystyle\sum_{<ij>}\rho_i\rho_j+0.93 \sum_{<<ij>>}\rho_i\rho_j+0.93\sum_{ijkl\in \square}c^\dagger_i c^\dagger_k c_jc_l+0.65\sum_{ijk\in \Delta}\rho_i c^\dagger_j c_k  $\\ \hline
Dirac  &\ 1 &\ $\displaystyle\sum_{<ij>}\rho_{i2}\rho_{j2}+0.4 \sum_{<<ij>>}\rho_{i2}\rho_{j2} +0.32 \sum_i \rho_{i1}\rho_{i2} -0.27 \sum_{<<<<ij>>>>}\rho_{i2}\rho_{j2}  $ \\   \hline
D-wave  &\ 2 &\ $\displaystyle\sum_{<<ij>>}\rho_{i1}\rho_{j2} +0.76 \sum_{<ij>} \rho_{i1}\rho_{j2} +[0.57e^{2.5i}\sum_{ijk\in \Delta}\rho_{i2}c^\dagger_{j1}c_{k2}+h.c.]+0.44 \sum_{<<<ij>>>}\rho_{i2}\rho_{j2} -0.26\sum_{[ij]} \rho_{i2}\rho_{j2}$\\ \hline
Triangular &\ 2 &\ $\displaystyle\sum_{<<AB>>}\rho_A\rho_B +0.8 \sum_{ABC\in \Gamma}\rho_C c^\dagger_A c_B  $ \\ \hline
\end{tabular}
\caption{The largest contributions to the first pseudopotential $U^1$ for various models in the lattice basis. Here $\rho_{i\sigma}=c^\dagger_{i\sigma} c_{i\sigma}$ where the $i$ is the position index and $\sigma=1,2$ is the spin index, if any. The brackets $<ij>$ refers to nearest-neighbor (NN) sites, $<<ij>>$ to next-nearest-neighbor (NNN) sites, etc. while $[ij]$ is the shorthand for eighth neighbor (NNNNNNNN) sites (see Fig. \ref{triangular}). The symbols $\square$,$\Delta$ and $\Gamma$ refers to special configurational patterns described in the main text. Note that $A,B,C$ are inequivalent sites in the triangular lattice model. }
\label{table1}
\end{table}
\end{widetext}

\subsection{Checkerboard model}

\begin{figure}
\begin{minipage}{0.99\linewidth}
\includegraphics[width=0.98\linewidth]{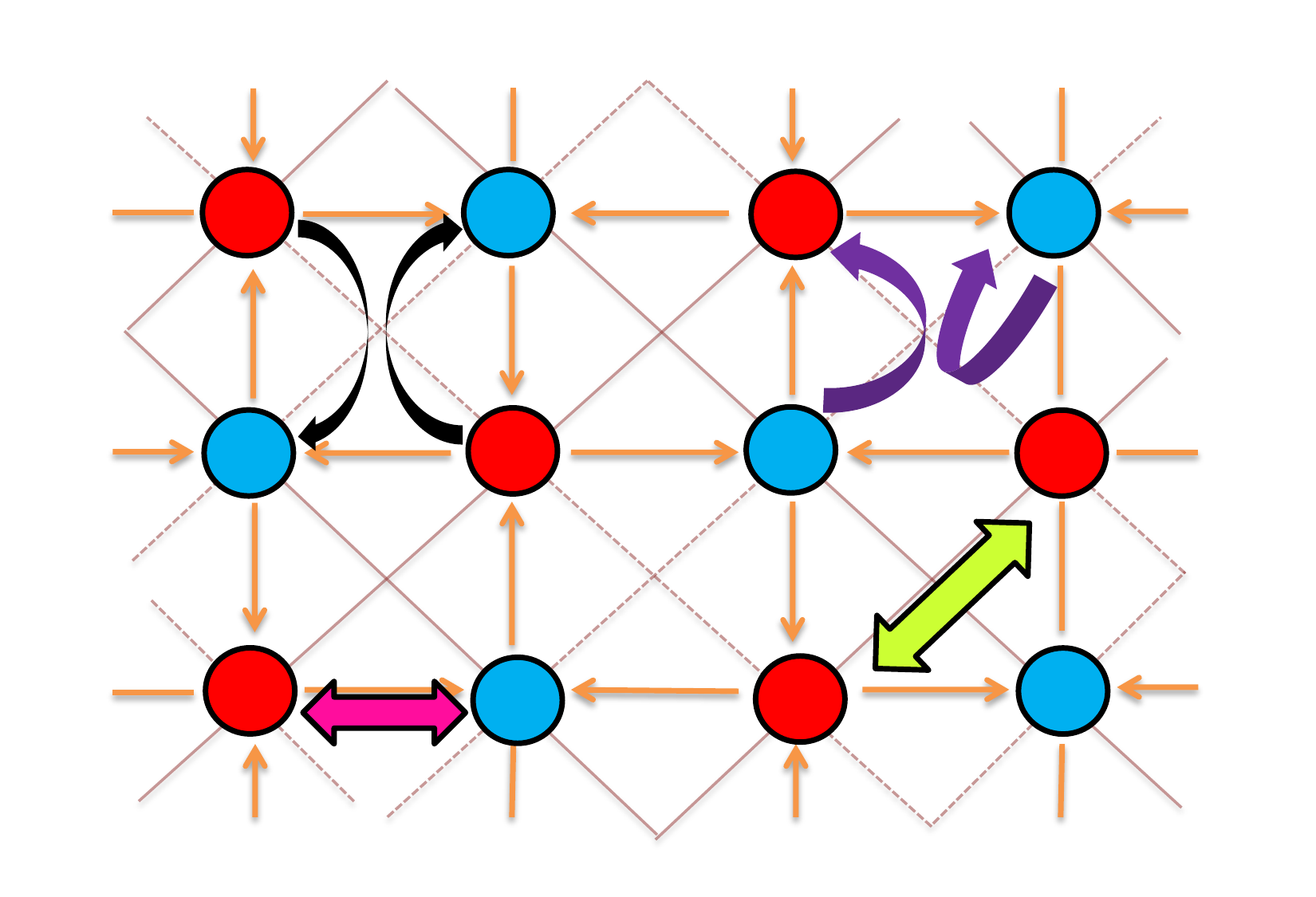}
\end{minipage}
\caption{(Color online) The truncated $U^1$ PP consists of two density-density interaction terms (pink and bright green of relative strengths $1$ and $0.93$) and two pair hopping terms (black and purple of strengths $0.93$ and $0.65$), as indicated in Eq. \ref{cbeq}. Density-density terms are represented by double-headed arrows. The black interaction involves two hoppings around a square plaquette. The purple interaction involves one self-hopping (i.e. density term) and one hopping among sites arranged in a triangle as shown. The red and blue sites sit on different sublattices. The thin arrows indicate the sign of the phase $\phi$ for the NN hoppings while the solid(dashed) diagonal lines indicate the sign of the NNN hoppings.}
\label{cb}
\end{figure}

The checkerboard (CB) model is a $C_1=1$ flat-band lattice model proposed in Ref. \onlinecite{Sun2011}. As shown in Fig.~\ref{cb}, it consists of two square lattices interlocked in a checkerboard fashion. The NN interactions are parametrized by $t$ and exists between sites belonging to different sublattices. They carry the phase $\phi\neq \pi$ that produces the time-reversal symmetry breaking necessary for a nonzero Chern number. The NNN hoppings take values of $t'$ or $-t'$ depending on whether they are connected by a solid or dashed line as shown in Fig. \ref{cb}. All in all, the hamiltonian is given by\cite{Lee2013}

\[H^{\text{CB}}(k)=d_0 I + \sum_i d_i \sigma_i, \]
where
\[ d_1= -2t[\cos\phi + \cos(k_x+k_y-\phi)+\cos(k_x+\phi)+\cos(k_y+\phi)], \]
\[ d_2= -2t[-\sin\phi + \sin(k_x+k_y-\phi)+\sin(k_x+\phi)+\sin(k_y+\phi)], \]
\[ d_3= -2t'(\cos k_x-\cos k_y).\]
The explicit expression for $d_0$ is omitted because it is not needed for the computation of the coherent state basis.

We set the parameters to be $t=1$, $t'=1/(2+\sqrt{2})$ and $\phi=\pi/4$ as in
 Ref. \onlinecite{Sun2011} to maximize the flatness of the
 occupied band. The dominant terms of $U^1$ are given by

\begin{eqnarray}
H&=&\displaystyle\sum_{<ij>}\rho_i\rho_j+0.93 \sum_{<<ij>>}\rho_i\rho_j+0.93\sum_{ijkl\in \square}c^\dagger_i c^\dagger_k c_jc_l\nonumber\\
& &+0.65\sum_{ijk\in \Delta}\rho_i c^\dagger_j c_k
\label{cbeq}
\end{eqnarray}
which consists of two density-density interaction ($\rho_i\rho_j$) terms and two pair hopping terms ($c^\dagger_i c^\dagger_k c_jc_l$ and $\rho_i c^\dagger_j c_k =c^\dagger_ic_i c^\dagger_j c_k $), as is illustrated in Fig. \ref{cb}.

\subsection{Dirac model}

The lattice Dirac model\cite{Qi2006} provides one of the simplest realizations of a $C_1=1$ system on a lattice. Each site admits a spin-$1/2$ degree of freedom, and is connected to each other only through NN hoppings. In momentum space, the model is defined as
\begin{equation} H^{\text{Dirac}}(k)=d_0 I + \sum_i d_i \sigma_i \label{dirac1}\end{equation}
with $d_1=\sin k_x -\sin k_y$, $d_2=\sin k_x+ \sin k_y$ and $d_3=m+\cos k_x+\cos k_y$. For $0 < \pm m< 2 $, the Dirac cones at $(k_x,k_y)=(n_x\pi,n_y\pi)$, $n_x,n_y\in \mathbb{Z}$ lead to a Chern number of $\pm1$ for the lower band. The band can be made approximately flat by adjusting $d_0$. 
It should be noted that the $d$-vector here has been chosen slightly differently from Ref. \onlinecite{Qi2006} for later convenience, but the two choices are equivalent by a simple basis rotation. The leading contributions of $U^1$ are given by
\begin{eqnarray}
H&=&\displaystyle\sum_{<ij>}\rho_{i2}\rho_{j2}+0.4 \sum_{<<ij>>}\rho_{i2}\rho_{j2}\notag \\
&& +0.32 \sum_i \rho_{i1}\rho_{i2} -0.27 \sum_{<<<<ij>>>>}\rho_{i2}\rho_{j2}\\
\label{diraceq}
\end{eqnarray}

\begin{figure}
\begin{minipage}{0.99\linewidth}
\includegraphics[width=0.99\linewidth]{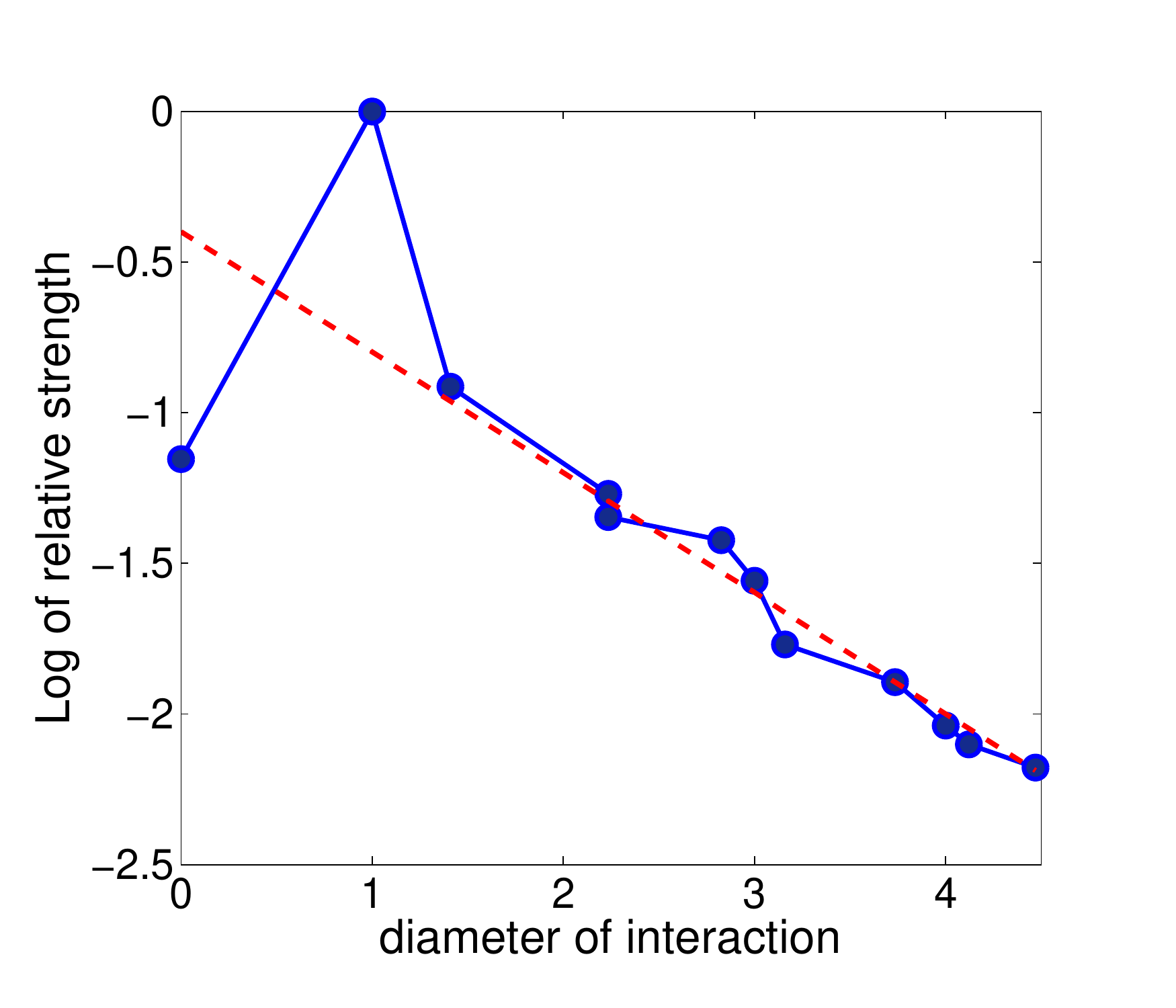}
\end{minipage}
\caption{(Color online) A Log plot of the relative strengths of the larger contributions to $U^1$ as a function of the diameter of the interaction contribution, i.e. spatial distance between the furthest pair in $c^\dagger_ic^\dagger_jc_kc_l$. The dominant term, which is normalized to have unit magnitude, is the NN density-density term with a diameter of $1$, while the term with smallest diameter is that of the same-site unequal spin density-density interaction. More than one type of interaction may have the same diameter, i.e. at diameter $\sqrt{5}\approx 2.23$. We observe an approximate spatial exponential decay of the interactions beyond the first few terms. The decay rate of $\approx 0.4$ depends on the model being studied.
}
\label{dirac}
\end{figure}

From this equation we can see that the $U^1$ PP is dominated by the first term (see Fig. \ref{dirac} (a)) which is a nearest neighbor density-density interaction between electrons with down spin or pseudospin. Although in Eq. (\ref{diraceq}) we have also kept other subleading terms, it is reasonable to conjecture that only the nearest neighbor term is already a good approximation to the PP Hamiltonian, which admits a Laughlin ground state at $1/3$ filling.

\begin{figure}
\begin{minipage}{0.99\linewidth}
\includegraphics[width=0.8\linewidth]{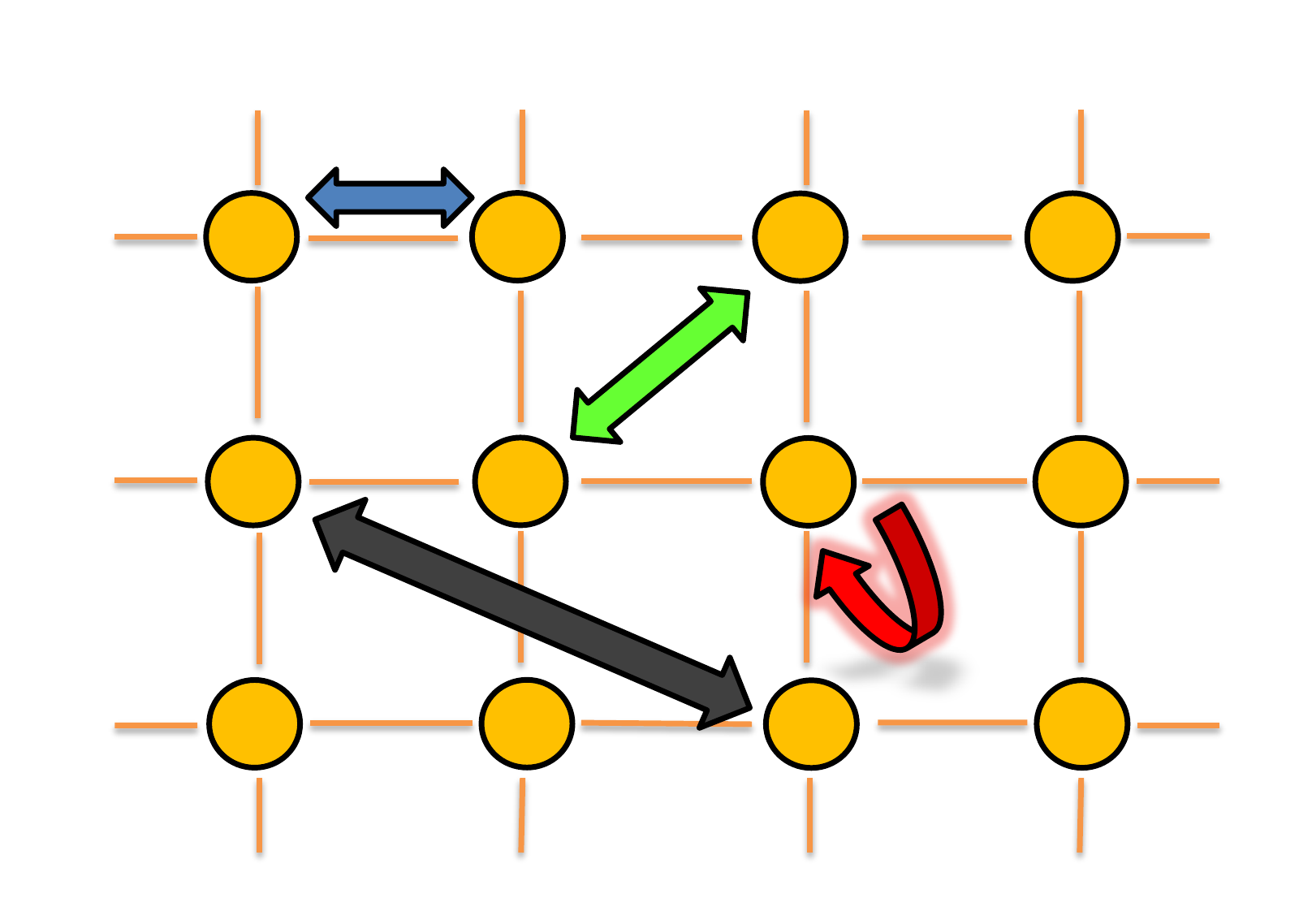}
\end{minipage}
\caption{(Color online) The truncated $U^1$ PP of the Dirac Model consists of four density-density interaction terms (blue, bright green,red and grey corresponding to relative strengths of $1$,$0.4$,$0.32$ and $0.27$), as indicated in Eq. \ref{diraceq}. The red arrow represents a density-density interaction between the up and down spins on the same site, while the other arrows all involve the down spins only.}
\label{diracmodel}
\end{figure}

\subsection{The Chern number $2$ D-wave model and its interpolation to two decoupled Dirac models}

Here we proceed to models with Chern number $2$, where the FCI system is mapped onto a bilayer QH system with decoupled layers. We consider a natural generalization of the Dirac model:
\begin{eqnarray}
H^d(k)&=&\sigma_x\left(\cos k_x-\cos k_y\right)+\sigma_z\left[\cos(k_x-k_y)\right.\nonumber\\
& &\left.-\cos(k_x+k_y)\right]+\sigma_y\left(\cos k_x+\cos k_y\right)\label{Hdwave}
\end{eqnarray}
This model can be viewed as a $d$-wave version of the lattice Dirac model, which is obtained by replacing the spin-dependent hopping terms $\sin k_x, \sin k_y$ in the Dirac model by the terms $\cos k_x-\cos k_y$ and $\cos(k_x-k_y)-\cos (k_x+k_y)$ with $d$-wave symmetry. The $d$-wave symmetry can be seen clearly by expanding this model near $(k_x,k_y)=(0,0)$, which leads to
\begin{eqnarray}
H^d(k)&\simeq &-\frac12\left(k_x^2-k_y^2\right)\sigma_x+2k_xk_y\sigma_z+2\sigma_y
\end{eqnarray}
The model has a Chern number $2$ for its lower band, which can be understood as coming from two quadratic touching points at $(0,0)$ and $(\pi,\pi)$ with the degeneracy lifted by the $\sigma_y$ term\cite{Chong2008,Sun2009}.

As is discussed in Sec. \ref{sec:preliminaries}, the Chern number $2$ model (on an even-by-even sized lattice) is mapped to two decoupled ``layers" of Landau levels. Two sets of coherent states can be constructed for these two layers, and pseudopotential Hamiltonians can be defined for layer-decoupled FCI states, {\it i.e.}, direct product of Laughlin states in each layer. For more generic states such as the Halperin $(mnl)$ states\cite{Halperin1983} with interlayer coupling, the pseudopotential Hamiltonian is not known, except the special case of $n=m,~l=n-1$ which is a singlet state\cite{Halperin1983}. Here we will focus on the first PP Hamiltonian for each layer, which has $(330)$ state as its ground state. This state is of particular interest since it is a topological nematic state\cite{Barkeshli2012} in which lattice dislocations become non-Abelian defects, and has not been realized in any existing FCI models.

Before computing the pseudopotential Hamiltonian for this model, it is interesting to note a relation of this model to the Dirac model discussed in the last subsection. Since Chern number is the only topological invariant for a 2D energy band, a model with a $C_1=2$ band can always be adiabatically deformed into one with two decoupled $C_1=1$ bands, as is demonstrated by the Wannier state mapping. For the model in Eq.(\ref{Hdwave}), this equivalence can also be shown explicitly by an adiabatic deformation of the Hamiltonian to that of two decoupled copies of Dirac model in Eq.(\ref{diracmodel}). For this purpose, consider the following parameterized Hamiltonian

\begin{figure}[H]
\begin{minipage}{\linewidth}
\includegraphics[width=.32\linewidth]{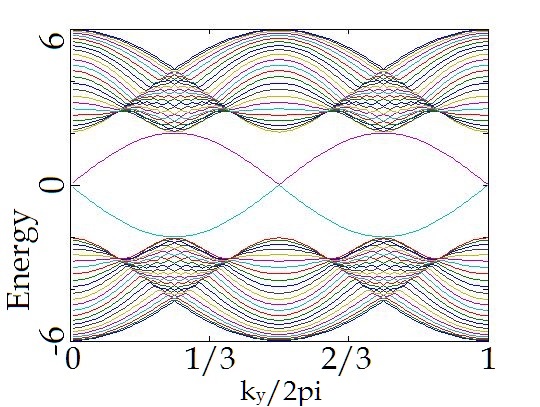}
\includegraphics[width=.32\linewidth]{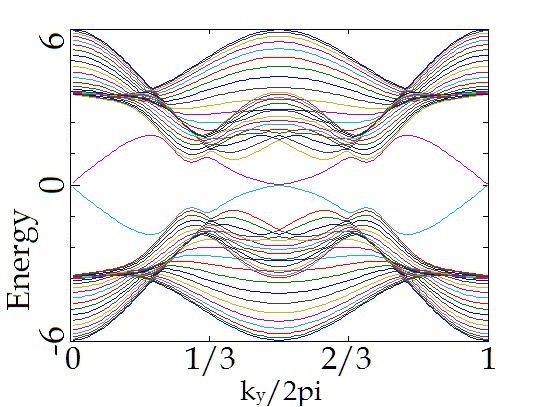}
\includegraphics[width=.32\linewidth]{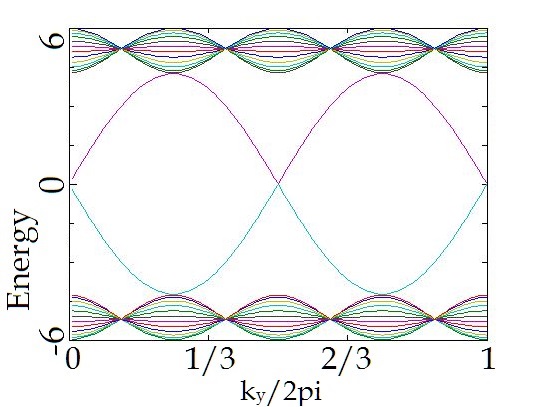}
\end{minipage}
\caption{Left to Right) The bandstructures of $H(0)$, $H(0.5)$ and $H(1)$ as a function of $k_y$. The edge states remains qualitatively the same during the entire interpolation, with no gap closure.}\label{interbands}\end{figure}

\begin{eqnarray}
 H_\lambda(k)&=&\sigma_x \left[(1-\lambda)[\sin (k_x+k_y)-\sin(k_x-k_y)]\right.\nonumber\\
 & &\left.+\lambda (\sin k_x -\sin k_y)\right]\nonumber\\
 && +\sigma_y\left[(1-\lambda)[\sin (k_x+k_y)+\sin(k_x-k_y)]\right.\nonumber\\
 & &\left.+\lambda (\sin k_x +\sin k_y)\right] \nonumber\\
 &&+ \sigma_z[(1-\lambda)+\cos(k_x+k_y)+\cos(k_x-k_y)]\nonumber \\
\label{interpol1}
\end{eqnarray}
$H_1(k)$ is equivalent to $H^d$ in Eq. (\ref{Hdwave}) by a translation $(k_x,k_y)\rightarrow \left(k_x+\frac{\pi}2,k_y+\frac{\pi}2\right)$.
$H_0(k)$ only contains second neighbor couplings, so that the two sublattices are decoupled, as is illustrated in Fig. \ref{interpol1} (Left panel). Restricted to each sublattice, this model is actually the Dirac model Eq. (\ref{dirac1}) with $m=1$. For all $\lambda\in[0,1]$, the model is gapped with $C_1=2$ in the occupied band, as is shown in Fig. \ref{interbands} by the energy spectrum on a cylindrical geometry. 
%

If we construct the Wannier states of this model in the ordinary way by Fourier transforming the Bloch state along $k_x$ direction with fixed $k_y$ at $\lambda=0$, the Wannier states and their correspondingly coherent states do not have a direct relation with those of the Dirac model since the translation along $x$ direction exchanges the two sublattices. From the last subsection, we have observed that the PP Hamiltonian of the lattice Dirac model has a nice short-ranged form. We will thus like to define an alternative Wannier basis for the $C=2$ model which is localized along the diagonal $(1,1)$ direction.

\begin{figure}
\begin{minipage}{0.99\linewidth}
\includegraphics[width=0.48\linewidth]{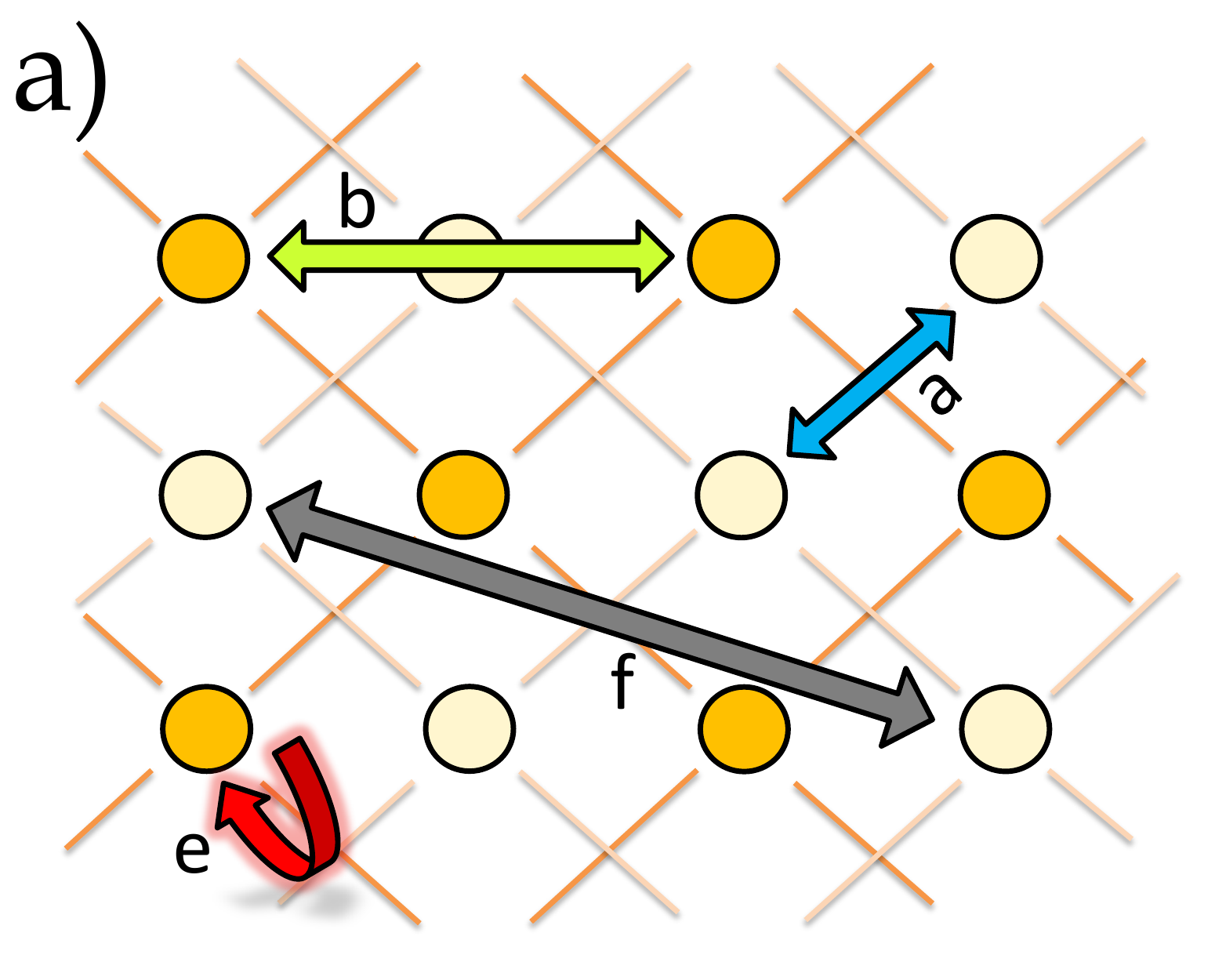}
\includegraphics[width=0.48\linewidth]{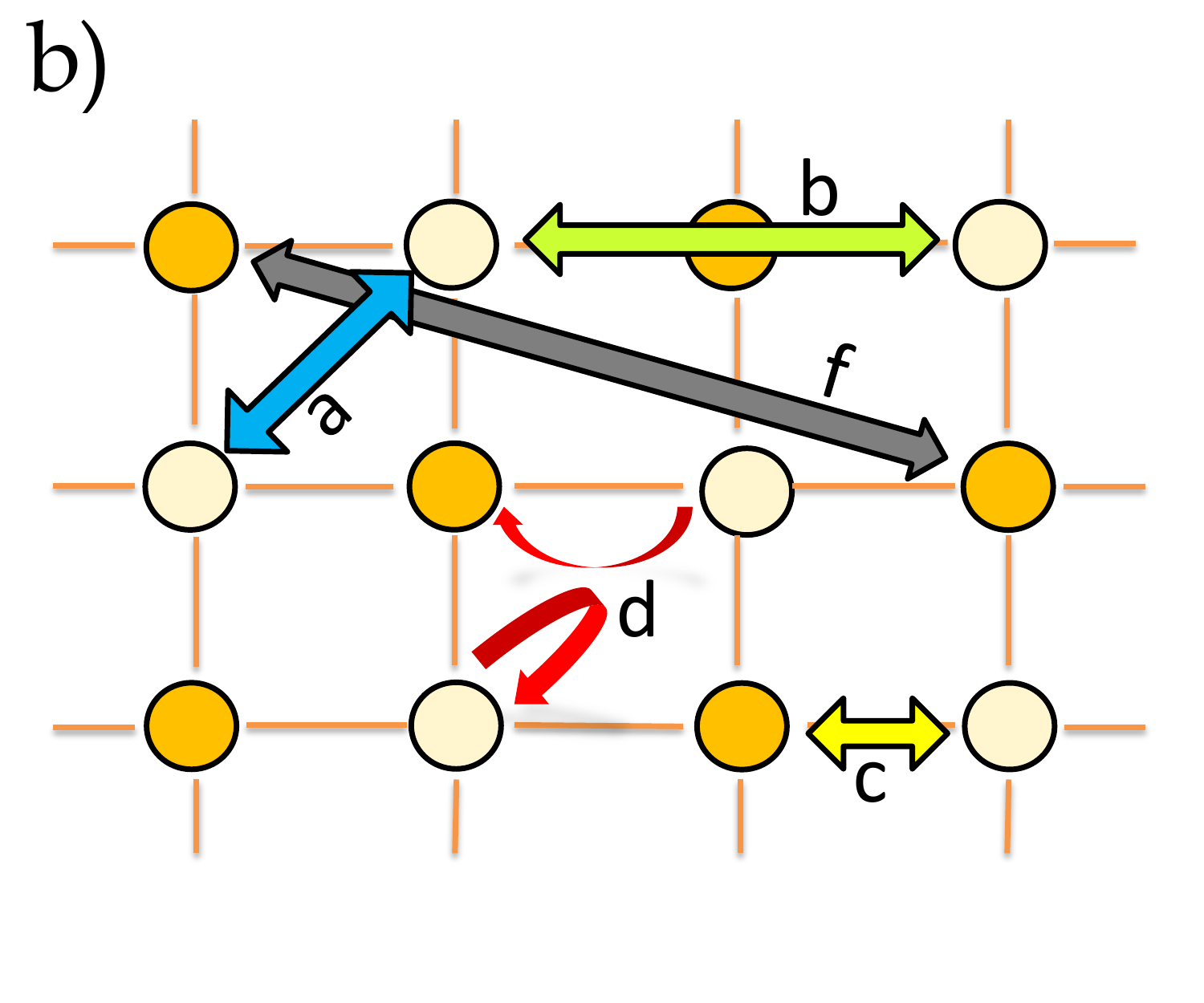}
\end{minipage}
\caption{(Color online) a) The truncated $U^1$ PP for the decoupled Dirac limit $H'_{\lambda=0}$. It has the same interactions as that of the original Dirac model (Fig. \ref{diracmodel}), but is rescaled by a factor of $\sqrt{2}$ and rotated by $\pi/4$. For instance, the NN interaction in the original Dirac model becomes the NNN interaction. Note that all interactions connect only sites in the same sublattice, as indicated by the diagonal bonds illustrated above.
b) The truncated $U^1$ PP for the d-wave limit $H'_{\lambda=1}$. It consists of four density-density interaction terms (blue, yellow light green and grey of relative strengths $1$, $0.76$, $0.44$ and $0.26$) and a pair hopping term (red of relative strength $0.57$), as indicated in Eq. \ref{dwaveeq}. The red interaction involves one self-hopping (i.e. density term) and one hopping among sites arranged in a triangle as shown. There is no decoupling between the sublattices in this case. All interactions are labeled by letters which will appear again in Fig. \ref{strengthgraph}.}
\label{interpol}
\end{figure}

The diagonally directed Wannier states and coherent states can be constructed through a redefinition of the unit cell $x'=x-y,y'=y$ or $k'_x=k_x,k'_y=k_x+k_y$. The Wannier states defined after the unit cell redefinition are $k_y'$ eigenstates. A flux that couples to $k'_y$ will now cause a spectral flow of the WFs in the $x'$ direction which thus stays in the same sublattice. In this definition, the decomposition of the $C_1=2$ band into two layers automatically reduce to the decomposition to the Dirac model on the two sublattices in the $\lambda=0$ limit. In this new coordinates, the hamiltonian reads
\begin{eqnarray}
H'_\lambda(k') &=&\sigma_x \left[(1-\lambda)[\sin k'_y-\sin (2k'_x-k'_y)]\right.\nonumber\\
 & &\left.+\lambda (\sin k'_x -\sin (k'_y-k'_x))\right]\notag\\
 && +\sigma_y \left[(1-\lambda)[\sin k'_y+\sin(2k'_x-k'_y)]\right.\nonumber\\
 & &\left.+\lambda (\sin k'_x +\sin (k'_y-k'_x))\right] \notag\\
 &&+ \sigma_z\left[(1-\lambda)+\cos k'_y+\cos(2k'_x-k'_y)\right]
\label{interpol2}
\end{eqnarray}
This is the expression of the hamiltonian which we will use for calculating $U^1$. For $\lambda=0$ the Hamiltonian reduces to that of the $C_1=1$ Dirac hamiltonian Eq. \ref{dirac1} if we transform its coordinates to $r''=(x'',y'')$ on each sublattice:
\begin{eqnarray}
x''&=&\frac{1}{2}x'=\frac{x-y}{2},~
y''=\frac{1}{2}x'+y'=\frac{x+y}{2}\nonumber\\
k''_x&=&2k'_x-k'_y,~
k''_y=k'_y
\end{eqnarray}
This unit cell redefinition belongs to the type described in Eqs. \ref{abredef1} and \ref{abredef2} with $\alpha=\frac{1}{2}$ and $\beta = -\frac{1}{2}$. As elucidated in the sentences preceding them, the coherent states in the new coordinates will be identical to those from the old ones as long as the periodic part of the Bloch state $\phi(k')$ remain invariant under such a transformation. Indeed, we mathematically see why the PP of the decoupled Dirac model ($\lambda=0$) is identical to that of the $C_1=1$ Dirac model up to rescaling and rotation.

\begin{figure}[H]
\begin{minipage}{\linewidth}
\includegraphics[width=.99\linewidth]{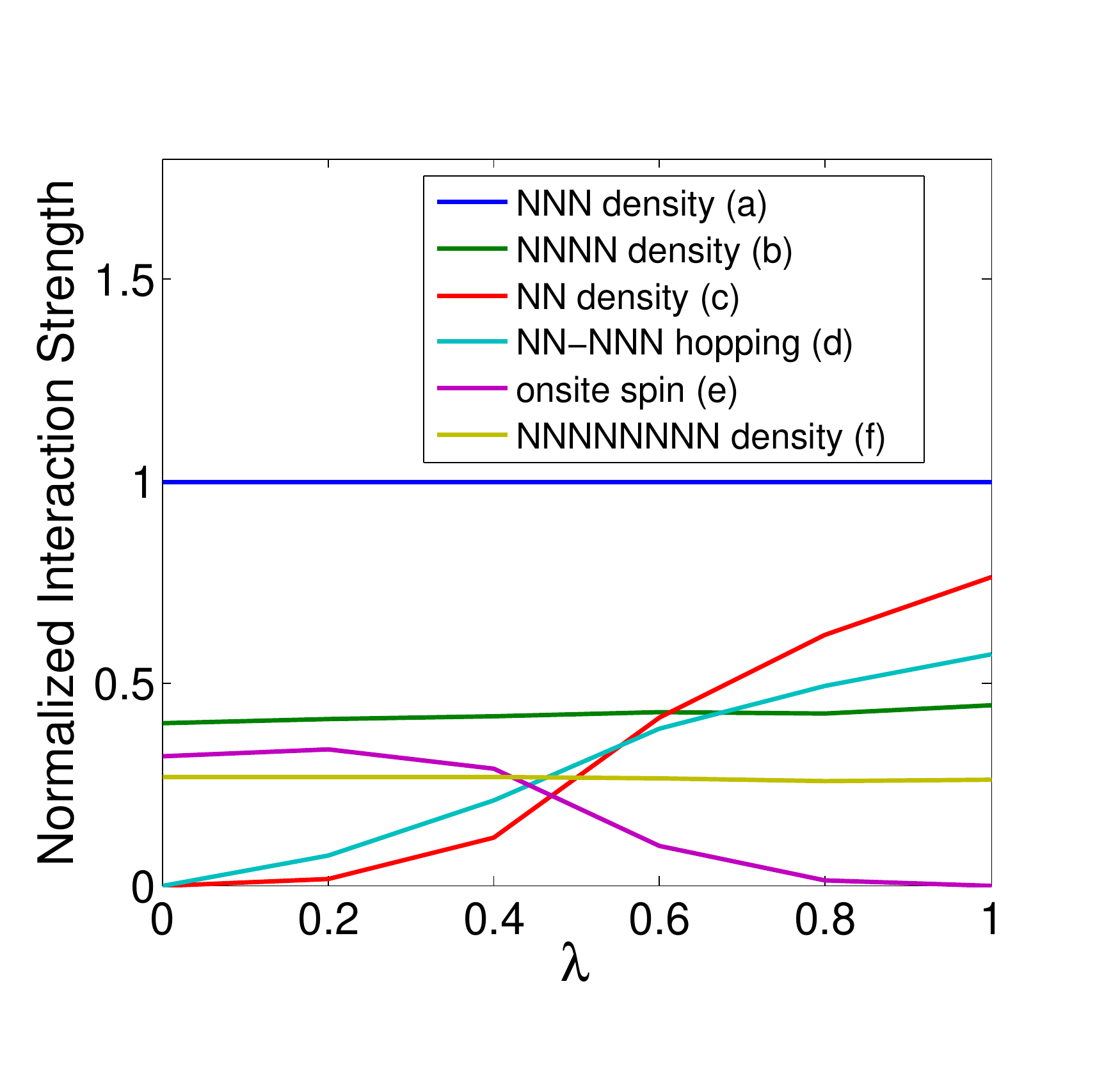}
\end{minipage}
\caption{The relative magnitudes of the various interaction terms in $U^1$ of $H'_\lambda$. The amplitude of the leading term has been normalized to $1$. Terms which are present in both models remain stable throughout the interpolation. Their exact definitions can be found in Fig. \ref{interpol}, where they are labelled graphically from a) to f). }
\label{strengthgraph}\end{figure}

The numerical results of the truncated PP Hamiltonian is shown in Fig. \ref{interpol} and \ref{strengthgraph}. For $\lambda=0$, the dominant terms of $U^1$ is identical to that in Eq. \ref{diraceq} after rescaling and $\frac{\pi}4$ rotation. For the $d$-wave model at $\lambda=1$, the PP Hamiltonian is given by
\begin{eqnarray}
H&=&\displaystyle\sum_{<<ij>>}\rho_{i1}\rho_{j2} +0.76 \sum_{<ij>} \rho_{i1}\rho_{j2}\notag\\
&& +[0.57e^{2.5i}\sum_{ijk\in \Delta}\rho_{i2}c^\dagger_{j1}c_{k2}+h.c.]\notag\\
&&+0.44 \sum_{<<<ij>>>}\rho_{i2}\rho_{j2} -0.26\sum_{[ij]} \rho_{i2}\rho_{j2}
\label{dwaveeq}
\end{eqnarray}
As shown in Fig. \ref{strengthgraph}, the leading term remains the same as $\lambda=0$, while the relative magnitudes of NNN, NNNN and NNNNNNNN (eighth-nearest neighbor) density-density interaction terms of $U^1$ remain stable across the interpolation. This attests to the robustness of our PP construction w.r.t. deformations that do not change the topology of the model. The other three terms vary significantly because they exist only in $\lambda\neq 0$ models. For instance, the NN density-density and NN-NNN hopping terms do not exist in the decoupled Dirac limit because the NN site does not belong to the same sublattice.

\subsection{Triangular lattice model}

\begin{figure}[H]
\begin{minipage}{0.99\linewidth}
\includegraphics[width=.5\linewidth]{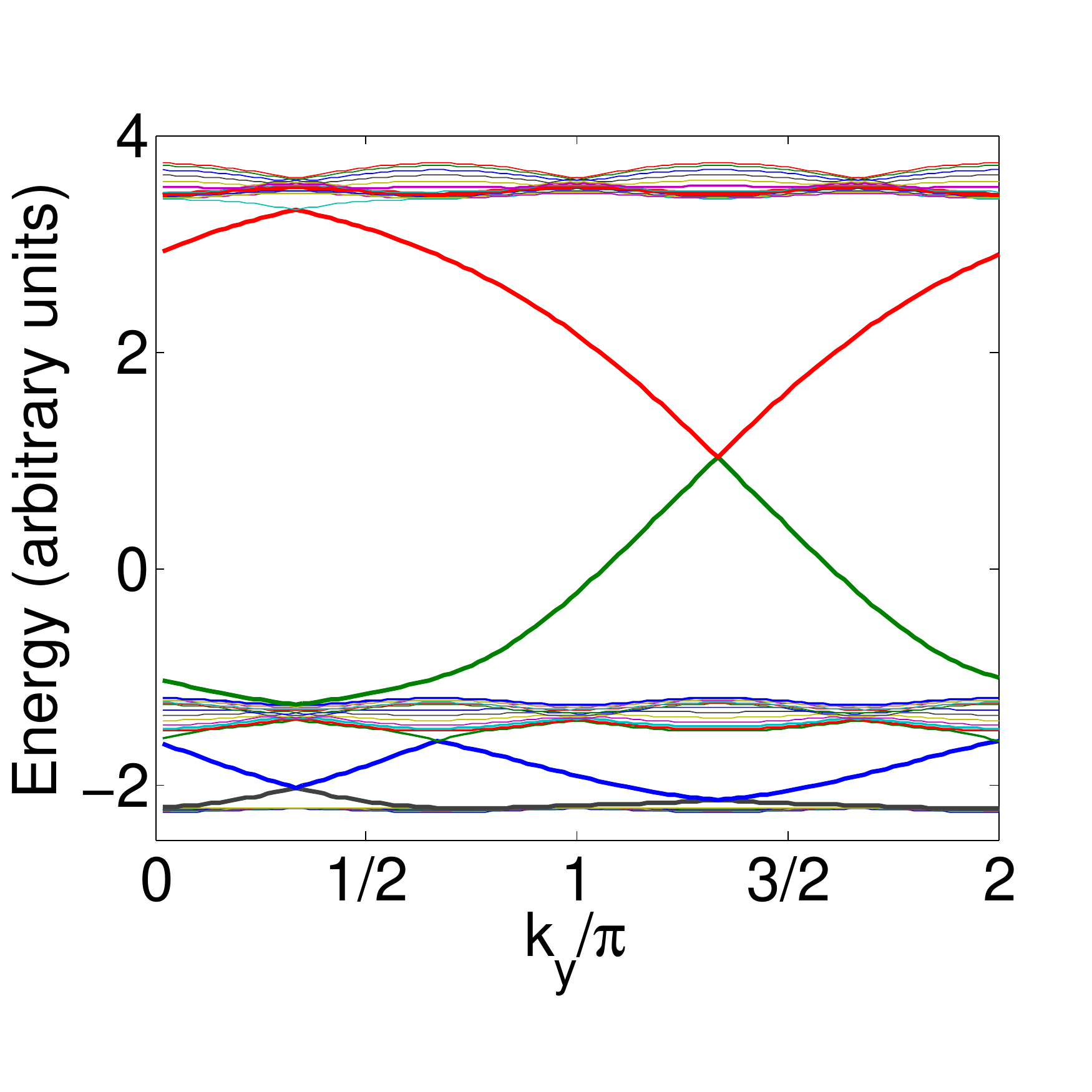}
\includegraphics[width=.5\linewidth]{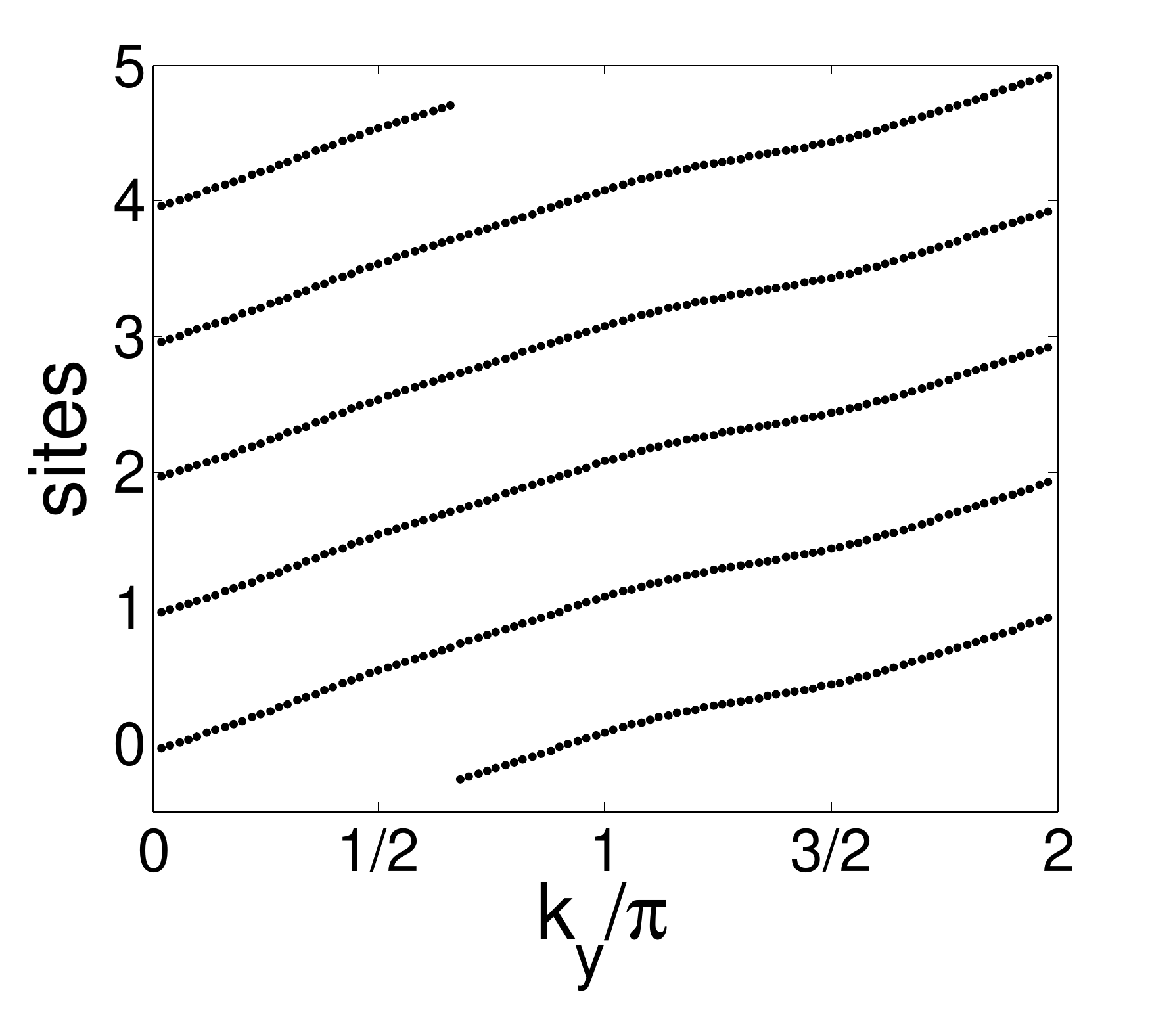}
\end{minipage}
\caption{Left) Band structure of the triangle lattice model, which is the same as in Ref. \onlinecite{Wang2011}. We can see the flatness of the lowest occupied band. Right) The Wannier polarization evolves by two sites as the flux $k_y\rightarrow k_y+A_y$ cycles over $2\pi$. }
\label{triangularband}
\end{figure}

As a last example, we consider another $C_1=2$ model, the 3-band triangular lattice flatband model introduced in Ref. \onlinecite{Wang2011}. Its lowest (occupied) topological flat band carries a Chern number of $2$, as evidenced in the presence of its two edge states on Fig. \ref{triangularband}. Each unit cell contains three inequivalent spinless sites, leading to three bands with asymmetrical dispersions. From its real-space description detailed in Ref. \onlinecite{Wang2011}, we obtain its momentum-space hamiltonian $h^{Tri}_{ij}(k)$ with
\begin{eqnarray}
h_{11}&=&t'(\cos(k_y+\phi)+\cos(k_x-\phi)+\cos(k_y-k_x-\phi))\notag\\
h_{12}&=& t(1+e^{i(k_x-k_y)}+e^{-ik_y})\notag\\
h_{13}&=& t(e^{i(k_x-2k_y)}+e^{i(k_x-k_y+2\phi)}+e^{-i(k_y+2\phi)})\notag\\
h_{22}&=&t'(\cos(k_y-\phi)+\cos(k_x+\phi)+\cos(k_y-k_x+\phi))\notag\\
h_{23}&=&-t(1+e^{i(k_x-k_y-2\phi)}+e^{i(2\phi-k_y)})\notag\\
h_{33}&=& -t'(\cos k_y+\cos k_x +\cos(k_y-k_x))
\end{eqnarray}
with $h_{ij}=h^*_{ji}$. $t$ and $t'$ parameterizes the magnitudes of the NN and NNN hoppings respectively, and $\phi$ provides the necessary time-reversal symmetry breaking. These parameters are chosen to take values of $t=t'=1/4$ and $\phi=\pi/3$ for maximum flatness of the occupied band.

\begin{figure}
\begin{minipage}{0.99\linewidth}
\includegraphics[width=0.8\linewidth]{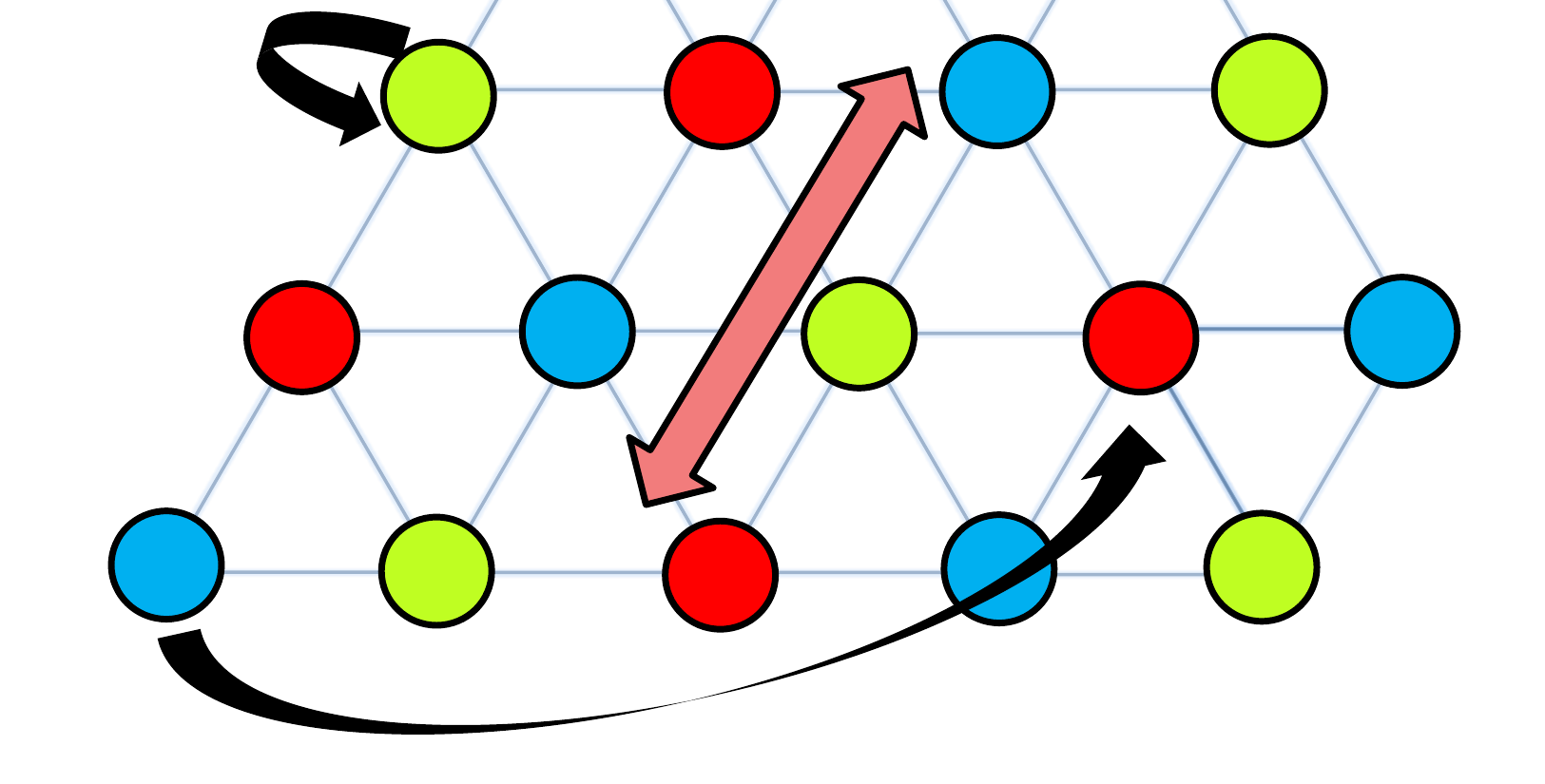}
\end{minipage}
\caption{(Color online) The truncated $U^1$ PP of the triangular model consists of a leading density-density interaction term (pink) followed by a pair  hopping term (black of relative strength $0.8$), as indicated in Eq. \ref{triangulareq}. As in previous figures, the density-density term is represented by a double-headed arrow. The black hoppings involve one self-hopping (density interaction) and a rather far jump. Note that the pink interaction between sites of type A,B is not equivalent to a similar-looking one between NNN A,C and B,C pairs. }
\label{triangular}
\end{figure}

For this model, the dominant terms in $U^1$ are given by
\begin{eqnarray}
H&=&\displaystyle\sum_{<<AB>>}\rho_A\rho_B +0.8 \sum_{ABC\in \Gamma}\rho_C c^\dagger_A c_B
\label{triangulareq}
\end{eqnarray}
where A,B and C refers to the inequivalent sites in the unit cell shown in Fig. \ref{triangular}. The notation $<<AB>>$, for instance, refers to the NNN A and B pairs, and not the NNN pairs among all three types.

In this model, the subleading interaction (of magnitude $0.8$) is still almost as large as the leading one. This is because the coherent states in this model are less localized than other models we studied. As captured in Eq. \ref{j}, the spread of the coherent states increases with the quantity $J$ which is a measure of the total complexity of the Bloch states. This complexity increases with the number of bands $N$. In addition, it is difficult to find a distinctive dominant term when $N$ is large because the number of possible interaction terms scales like $N^4$.

\section{Conclusion}
\label{sec:concl}

In this work, we have seen how the Wannier state representation can be used to "reverse engineer" a PP Hamiltonian in generic FCI systems. This enables us to once and for all write down the exact interaction Hamiltonian that admits certain desired groundstates. However, there is an inevitable proliferation of terms and some truncation in the range of interaction will be necessary in any practical calculation. The truncation error will be minimized if we use the Coulomb Gauge and choose an FQH basis corresponding to a special coherent state aspect ratio $A_{opt}=J/\pi+\pi/3$, where $J$ depends only on the geometric properties of the FCI system, namely its Berry curvature and Fubini-Study metric. To illustrate the robustness of our approach, we numerically obtained the first PP in various models with Chern number $1$ and $2$, including the CB model, Dirac model, d-wave model and triangular lattice model. The PPs are found to be reasonably well-approximated by a few dominating terms shown in Table I, especially for that of the flattened Dirac model where only one term dominates. Hence the central prediction of our work is that topological nematic states are likely to be realized with the interaction Hamiltonians shown in Table I (or perhaps even further truncated versions thereof).

As such, a natural follow-up to this work will be the numerical verification of topological nematic states in the $C=2$ models. If realized, this state will have non-abelian lattice dislocations with quantum dimension $\sqrt{3}$.
Although we have not performed explicit calculations with Chern numbers higher than $2$ (which are mapped to FQH multilayers) and/or higher PPs, such extensions can be achieved in a completely analogous way to what we have done. The challenge will be to find FCI systems where the proliferation of terms is sufficiently benign for a sensible truncation to still support their coveted exotic states.


\begin{acknowledgements}
We thank D.~N.~Sheng, Ronny Thomale and Yi Zhang for discussions. CH is supported by a scholarship from the Agency of Science, Technology and Research of Singapore.  XLQ is supported by the Packard foundation and the Alfred P. Sloan foundation.
\end{acknowledgements}

\appendix

\section{Pseudopotentials in terms of coherent states}
\label{ppcoherent}

In this appendix, we shall show how Eq. \ref{main}, where the PP $U^m$ is expressed in terms of the coherent state operators, simplifies to its usual definition in Eq. \ref{lag}. To start, it is first helpful to replace the gradient operators $\nabla_z^r$ by their conjugate variables $q$:

\begin{widetext}
\begin{eqnarray}
U^m &\propto& \int dz_1dz_2 \sum_r v^m_r l_B^{2r}\nabla_z^r (c^\dagger_z c_z ) \nabla_z^r (c^\dagger_z c_z )\notag\\
&\propto& \sum_r \int d^2q v^m_r q^{2r} \rho(-q)\rho(q)\notag\\
&=& \sum_r v^m_r \int d^2q l_B^{2r}q^{2r} \int d^2z e^{iq\cdot z} c^\dagger_z c_z \int d^2z' e^{-iq\cdot z'}c^\dagger_{z'}c_{z'}\notag\\
&\propto & \sum_r v^m_r  \int d^2q l_B^{2r}q^{2r} \int d^2z \int d^2z' e^{iq\cdot (z-z')} c^\dagger_z c^\dagger_{z'}c_{z'}c_z+\text{quadratic}
\end{eqnarray}
\end{widetext}

where $v^m_r$ are the coefficients of the $m^{th}$ Laguerre polynomial. Now we substitute the explicit expression of the coherent state operators from Eq. \ref{coherentdef}:

\begin{equation} c^\dagger_z =\sum_K e^{iz_2K}e^{-\pi A(z_1/C_1-K/(2\pi))^2}a^\dagger_K \end{equation}

where $a^\dagger_K$ creates the LLL Landau gauge wavefunction $\ket{\psi_K}$ given by Eq. \ref{LLLbasis}. With a slight abuse of notation of $z$ as being both a vector and complex number,

\begin{widetext}
\begin{eqnarray}
&&U^m \\
&\propto& \sum_r \sum_{K_1...K_4}  v^m_r \int d^2q l_B^{2r}q^{2r} \int d^2z \int d^2z' e^{iq\cdot (z-z')} e^{iz_2K_1}e^{-\pi A(z_1/C_1-K_1/(2\pi))^2}\notag\\
&&e^{iz'_2K_2}e^{-\pi A(z'_1/C_1-K_2/(2\pi))^2}e^{-iz'_2K_3}e^{-\pi A(z'_1/C_1-K_3/(2\pi))^2}e^{-iz_2K_4}e^{-\pi A(z_1/C_1-K_4/(2\pi))^2}a^\dagger_{K_1} a^\dagger_{K_2} a_{K_3}a_{K_4} \notag\\
&\propto& \sum_r \sum_{K_1...K_4} v^m_r \int d^2q l_B^{2r}q^{2r} \int d^2z \int d^2z' e^{i(K_1-K_4)z_2}e^{i(K_2-K_3)z'_2}e^{iq\cdot (z-z')} e^{\frac{A}{C_1}(z_1(K_1+K_4)+z'_1(K_2+K_3))}\notag\\
&& e^{-A(x^2+x'^2)/(C_1^2l_B^2)}e^{-\frac{A}{4\pi}\sum_i K_i^2}a^\dagger_{K_1}a^\dagger_{K_2} a_{K_3}a_{K_4}\notag\\
&\propto& \sum_r \sum_{K_1...K_4} v^m_r \int d^2q l_B^{2r}q^{2r} \int d^2z \int d^2z' e^{iq\cdot (z-z')} e^{iK_1z_2-\frac{A}{2C_1}(z_1/l_B-l_B K_1)^2}e^{iK_2z'_2-\frac{A}{2C_1}(z'_1/l_B-l_B K_2)^2}\notag\\
&& e^{-iK_3z'_2-\frac{A}{2C_1}(z'_1/l_B-l_B K_3)^2}e^{-iK_4z_2-\frac{A}{2C_1}(z_1/l_B-l_B K_4)^2}a^\dagger_{K_1}a^\dagger_{K_2} a_{K_3}a_{K_4}\notag\\
&\propto &\frac{V_0}{4}\sum_{K_1...K_4}\int \frac{d^2q}{(2\pi)^2}\int d^2z d^2z' L_m(q^2l_B^2)e^{i q\cdot (z-z')}\psi^*_{K_1}(z)\psi^*_{K_2}(z')\psi_{K_3}(z')\psi_{K_4}(z)a^\dagger_{K_1}a^\dagger_{K_2} a_{K_3}a_{K_4} \notag\\
&=&\frac{V_0}{4\sqrt{2\pi l_B^2}}\int d^2r d^2r'\psi^\dagger(r)\psi^\dagger(r') L_m(l_B^2 \nabla^2)(\delta(r-r'))\psi(r')\psi(r)
\end{eqnarray}
\end{widetext}

where

\[\psi_{K=\frac{2\pi n }{L}}(r)=\frac{1}{\sqrt{\sqrt{\pi}Ll_B}}e^{-iKy}e^{-\frac{A}{2C_1}(\frac{x}{l_B}-l_BK)^2}\]

is the LLL Landau gauge wavefunction.

For $m=1$, the $r=0$ term vanishes due to fermionic antisymmetry. The $(\nabla_z c_z^\dagger\cdot \nabla_z c_z^\dagger) c_zc_z$ and $(\nabla_z c_z\cdot \nabla_z c_z) c_z^\dagger c_z^\dagger$ parts of the $r=1$ term also vanishes due to the same reason. Hence we are left with \[V^1 \propto \int dz_1dz_2 c_z(\nabla_z c_z^\dagger\cdot \nabla_z c_z)c_z^\dagger \propto \int dz_1dz_2 c_z^\dagger \nabla_z c_z^\dagger c_z \nabla_z c_z\] which is the same manifestly local expression in \cite{Qi2011}. The unimportant proportionality constant has the same units for all $m$, and can be deduced from dimensional analysis and normalization.

\section{Derivation of the Coulomb Gauge condition}
\label{gaugederivation}

We perform the Euler-Lagrange minimization $\frac{\delta I}{\delta \theta}=0$ on $I[\theta]$ given by Eq. \ref{coherentI}. First, note that $I$ only depends explicitly on $\nabla_k \theta$, not $\theta$, because $|\nabla_k(e^{i\theta }f)|^2=|if\nabla_k\theta+\nabla f |^2$ where $f$ represents the remaining part of the integrand in $I[\theta]$ not containing $\theta$. Hence

\begin{widetext}
\begin{eqnarray}
0&=&\frac{\delta I}{\delta \theta}=\nabla_k \cdot \frac{\partial I}{\partial (\nabla_k \theta)}\notag\\
&=&  \sum_m \int_{[-1/2,1/2]^2}d^2z \nabla_k\cdot (-i (e^{-E_m+ik_yz_2}\phi)^\dagger \nabla_k (e^{-E_m+ik_yz_2}\phi)+e^{-2E_m}\nabla_k \theta)\nonumber \\
&=&\sum_m\int_{-1/2}^{1/2} dz_1\int_{-1/2}^{1/2} dz_2 e^{-2E_m}\left(-2\partial_y E_m(z_2+i\partial_y E_m+\partial_y\theta)+\nabla_k\cdot \vec{a}+a_y(iz_2-\partial_y E_m) +\frac{iA}{2\pi}+\nabla_k^2 \theta\right)  \nonumber \\
&=& \int_{-1/2}^{1/2} dz_2 \int_{-\infty}^{\infty}d\eta e^{-2\pi A\eta^2}\left(-2A\eta(z_2+iA\eta+\partial_y\theta)+\nabla_k\cdot\vec{a}+a_y(iz_2-\eta A) +\frac{iA}{2\pi}+\nabla_k^2 \theta\right)  \nonumber \\
&=& \int_{-1/2}^{1/2} dz_2 \int_{-\infty}^{\infty}d\eta e^{-2\pi A\eta^2}\left(-2iA^2\eta^2+\nabla_k\cdot\vec{a} +\frac{iA}{2\pi}+\nabla_k^2 \theta\right)  \nonumber \\
&=& \int_{-1/2}^{1/2} dz_2 \frac{1}{\sqrt{2}}\left(\nabla_k\cdot\vec{a} +\nabla_k^2 \theta+0\right)  \nonumber \\
&\propto & \nabla_k\cdot\vec{a} +\nabla_k^2 \theta \notag \\
&=&\nabla_k\cdot(\vec{a}+\nabla_k \theta)
\label{eq:el}
\end{eqnarray}
\end{widetext}

Some steps deserve explanation. I have defined $E_m=-\pi A \left(\frac{k_y}{2\pi}-\frac{z_1}{C_1}+m\right )^2=-\pi A \eta^2$ in line $3$, so that $\partial_y E_m = A\eta$. This combines $z_1$ and $m$ into one continuous variable $\eta$, thereby reducing the infinite sum into one simple gaussian integral. We also note that the factor multiplying $a_y$ in the third line disappears because it is linear in $\eta$ and $z_2$. This allows the final expression to be symmetric in $a_x$ and $a_y$.

The terms involving $A$ cancels nicely, implying that the optimal phase is independent of the aspect ratio $A$ used for the gaussian envelope. However there still exists a certain $A$ that will gives the minimal spread in the coherent state, and that will be derived in the subsection after the next.

Actually, the MLWF phase $\theta_W$ can also be derived via a conceptually identical E-L minimization approach\cite{Kivelson1982}. If we do not integrate over $k_y$ or sum over $m$, do not involve $z$ (i.e. just let $A=0$) and treat $k_y$ as a parameter, the minimal $\langle r^2\rangle $ will still be given by

\[ \nabla_k\cdot(\vec{a}+\nabla_k \theta)\]

This is obvious from the derivation of Eq.~\ref{eq:el}. WLOG, let us choose to work in gauges where $a_y=0$. Noting that $\nabla_k=\partial_{k_x}$ when $k_y$ is a parameter,

\begin{equation}\nabla_k^2 \theta = \partial^2_{k_x} \theta = -\partial_{k_x} a_x \end{equation}

The first term of $\theta_W$ (shown in Eq.~\ref{eq:XL}) solves this equation up to a term linear in $k_x$. The latter enforces the periodicity of $2\pi$ in $k_x$, and can thus be uniquely determined up to an irrelevant overall phase.

\section{The spread of the coherent state}
\label{spread}

Here is how the average spread of the coherent state can be computed:

\begin{widetext}
\begin{eqnarray}
I[A]&=&\int_{[-1/2,1/2]^2} d^2z\langle r^2\rangle_{\psi_z}\notag\\
&=&\int_{[-1/2,1/2]^2} d^2z\sum_m \int d^2k \left|\nabla_k \left (e^{-E_m-ik_yz_2}e^{i\theta (k_x,k_y)}\phi(k_x,k_y)\right )\right |^2 \nonumber \\
&=&\int_{[-1/2,1/2]}dz_2 \int d^2k\int_{-\infty}^{\infty}d\eta  \left|\nabla_k \left (e^{-E_m-ik_yz_2}e^{i\theta (k_x,k_y)}\phi(k_x,k_y)\right )\right |^2 \nonumber \\
&=&\int_{-\frac{1}{2}}^{\frac{1}{2}}dz_2 \int d^2k\int_{-\infty}^{\infty}d\eta e^{-2\pi A \eta^2}\left(|\nabla_k \phi|^2 + (A\eta)^2 +z_2^2 + |\nabla_k \theta|^2 +2z_2 \partial_y \theta +\left[-i\vec{a}\cdot ((-A\eta +iz_2)\hat{y}+i\nabla_k \theta)+c.c.\right]  \right) \nonumber \\
&=&\int_{-\frac{1}{2}}^{\frac{1}{2}}dz_2 \int d^2k\int_{-\infty}^{\infty}d\eta e^{-2\pi A \eta^2}\left(|\nabla_k \phi|^2 + (A\eta)^2 +z_2^2 + |\nabla_k \theta|^2 +2\vec{a}\cdot \nabla_k \theta  \right) \nonumber \\
&=&\int_{-\frac{1}{2}}^{\frac{1}{2}}dz_2 \int d^2k\int_{-\infty}^{\infty}d\eta e^{-2\pi A \eta^2}\left(|\nabla_k \phi|^2 + (A\eta)^2 +z_2^2 + \nabla_k\theta\cdot(\nabla_k\theta +2\vec{a})  \right) \nonumber \\
&=&\int_{-\frac{1}{2}}^{\frac{1}{2}}dz_2 \int d^2k\int_{-\infty}^{\infty}d\eta e^{-2\pi A \eta^2}\left(|\nabla_k \phi|^2 + (A\eta)^2 +z_2^2 + |\vec{a}_{new}|^2-|\vec{a}|^2 \right) \nonumber \\
&=&\int_{-\frac{1}{2}}^{\frac{1}{2}}dz_2 \int d^2k \frac{1}{\sqrt{2A}}\left(|\nabla_k \phi|^2  +z_2^2 + |\vec{a}_{new}|^2-|\vec{a}|^2 +\frac{A}{4\pi}\right) \nonumber \\
&=&\frac{1}{\sqrt{2A}}\int d^2k \left(\left[|\nabla_k \phi|^2-|\vec{a}|^2\right]   + |\vec{a}_{new}|^2+\frac{1}{12} +\frac{A}{4\pi}\right) \nonumber \\
\label{eq:r2}
\end{eqnarray}
\end{widetext}
where, as before, $E_m=-\pi A \left(\frac{k_y}{2\pi}-\frac{z_1}{C_1}+m\right )^2=-\pi A \eta^2$.

\section{$\Omega_1$ in terms of projected position operators}
\label{omega1}

Following Ref. \onlinecite{Marzari1997}, we can also express $\Omega_1 $ as

\begin{eqnarray}
\Omega_1 &=& \int d^2k \left (|\nabla_k \phi|^2-|\vec{a}|^2\right ) \notag \\
&=& \int d^2k\left( \langle \nabla_k\phi|\nabla_k\phi \rangle - \langle \phi|\nabla_k \phi \rangle \langle \nabla_k \phi|\phi\rangle\right) \notag \\
&=& \int d^2k \langle \nabla_k \phi |Q|\nabla_k \phi \rangle\notag \\
&=& Tr(PxQx)+Tr(PyQy)\notag\\
&=& Tr(PxQ^2xP)+Tr(PyQ^2yP)\notag\\
&=&Tr[(PxQ)^2+(PyQ)^2]
\end{eqnarray}

where $P=|\phi\rangle\langle \phi|$ and $Q=I-P$. The sum of occupied bands is implied. The final expression is obviously a gauge-invariant quantity. The physical interpretation of $\Omega_1$ becomes clearer if we write

\begin{eqnarray}
\Omega_1 &=& Tr(PxQx)+Tr(PyQy)\notag\\
&=& Tr(Px^2) -Tr(PxPx)+ (x\leftrightarrow y) \notag\\
&=& Tr(P^2x^2)-Tr(P^2xP^2x) +(x\leftrightarrow y) \notag\\
&=& Tr(Px^2P)-Tr[(PxP)^2] +(x\leftrightarrow y) \notag\\
&=& Tr(Pr^2P)-Tr[\tilde{r}^2]
\end{eqnarray}

where $\tilde{x}=PxP$ is the operator whose eigenvectors are the MLWFs, and $\tilde{r}^2=\tilde{x}^2+\tilde{y}^2$. Hence $\Omega_1$ provides a certain measure of the spread of the state that exists within the subspace of occupied bands. This can be seen more clearly in the Wannier basis $|\vec R m\rangle=\frac{1}{4\pi^2}\int d^2k e^{-i\vec k \cdot \vec R} |\phi_m(\vec k )\rangle$, where\cite{Marzari1997}

\begin{equation} \Omega_1 = \sum_n \left[ \langle \vec 0n|r^2|\vec 0n\rangle -\sum_{\vec R m}|\langle \vec R m|\vec r|\vec 0 n\rangle|^2\right]
\end{equation}

We see that $\Omega_1$ is the mean-square spread of the WF (not necessarily maximally localized or even localized) minus a certain positive-definite quantity.

\bibliography{fci}


\end{document}